\documentclass[11pt]{article}

\def\Komabanumber#1{\hfill \begin{minipage}{4.2cm}\tt UT-Komaba #1
\end{minipage}}

\title{\vspace{20mm}\bf{Nonlinear gauge invariance and WZW-like action\\ for NS-NS superstring field theory}\vspace{25mm}}
\author{\Large{Hiroaki Matsunaga}\footnote{matsunaga@hep1.c.-tokyo.ac.jp} \vspace{3mm}}
\date{Institute of Physics, University of Tokyo \\ Komaba, Meguro-ku, Tokyo 153-8902, Japan \vspace{10mm}}

\usepackage{amsmath,amscd, amssymb}
\usepackage[hypertex]{hyperref}
\usepackage{cite}
\allowdisplaybreaks[3]

\makeatletter

  \@addtoreset{equation}{section}
\makeatother

\newcommand{\ld}{ [ \hspace{-0.6mm} [ }
\newcommand{\rd}{ ] \hspace{-0.6mm} ] }
\newcommand{\Ld}{ \big[ \hspace{-1.1mm} \big[ }
\newcommand{\Rd}{ \big] \hspace{-1.1mm} \big] }

\begin{document}

\maketitle 
{\vspace{-128mm} \Komabanumber{14-02}\vspace{128mm}}

\begin{abstract}
We complete the construction of a gauge-invariant action for NS-NS superstring field theory in the large Hilbert space begun in arXiv:1305.3893 by giving a closed-form expression for the action and nonlinear gauge transformations. % 33 words
The action has the WZW-like form and vertices are given by a pure-gauge solution of NS heterotic string field theory in the small Hilbert space of right movers. % 24 + 4 words
\end{abstract}

\thispagestyle{empty}
\clearpage

\tableofcontents
\setcounter{page}{1}

\section{Introduction}

While bosonic string field theories have been well-understood \cite{Witten:1985cc, Hata:1986, AlvarezGaume:1988bg, Sen:1990ff, Schubert:1991en, Zwiebach:1992ie, Gaberdiel:1997ia}, superstring field theories remain mysterious. 
A formulation of supersymmetric theories in the early days \cite{Witten:1986qs}, which is a natural extension of bosonic theory, has some disadvantages caused by picture-changing operators inserted into string products: singularities and broken gauge invariances \cite{Wendt:1987zh}. 
To remedy these, various approaches have been proposed within the same Hilbert space of $(\beta , \gamma )$ \cite{Arefeva:1989cp, Preitschopf:1989fc, AlvarezGaume:1988sj, Saroja:1992vw, Belopolsky:1997bg, Jurco:2013qra}. 

\vspace{2mm}

There exists an alternative formulation of superstring field theory: large space theory \cite{Berkovits:1995ab, Berkovits:1998bt, Berkovits:2004xh, Matsunaga:2013mba, Berkovits:2001im, Michishita:2004by, Kunitomo:2013mqa}. 
Large space theories are formulated by utilizing the extended Hilbert space of $(\xi , \eta , \phi )$ \cite{Friedan:1985ge} and the WZW-like action including no explicit insertions of picture-changing operators. 
One can check the variation of the action, the equation of motion, and gauge invariance without taking account of these operators. 
Of course, the action implicitly includes picture-changing operators, which appear when we concretely compute scattering amplitudes after gauge fixing. 
The singular behaivor of them is, however, completely regulated and there is no divergence \cite{Berkovits:1999bs,Iimori:2013kha}. 

\vspace{2mm}

The cancellation of singularities can also occur in the small Hilbert space. 
Recently, by the brilliant works of \cite{Erler:2013xta, Erler:2014eba}, it is revealed how to obtain gauge-invariant insertions of picture-changing operators into (super-) string products in the small Hilbert space: 
the NS and NS-NS sectors of superstring field theories in the small Hilbert space is completely formulated. 
%%%%%%%%%%%%%%%%%%%%%%%%%
In this paper, we find that using the elegant technique of \cite{Erler:2014eba}, one can construct the WZW-like action for NS-NS superstring field theory in the large Hilbert space. 

\vspace{2mm}

A pure-gauge solution of small-space theory is the key concept of WZW-like formulation of NS superstring field theory in the large Hilbert space, which determines the vertices of theory, and we expect that it goes in the case of the NS-NS sector. 
%%%%%%%%%%%%%%%%%%%%%%%%%
There is an attempt to construct non-vanishing interaction terms of NS-NS string fields utilizing a pure-gauge solution $\mathcal{G_{B}}$ of {\it bosonic} closed string field theory \cite{Matsunaga:2013mba}. 
However, the construction is not complete: the nonlinear gauge invariance is not clear and the defining equation of $\mathcal{G_{B}}$ is ambiguous. 
To obtain nonlinear gauge invariances, we have to add appropriate terms to these interaction terms defined by $\mathcal{G_{B}}$ at each order. 
Then, the ambiguities of vertices are removed and we obtain the defining equation of a suitable pure-gauge solution $\mathcal{G}_{L}$, which we explain in the following sections. 

\vspace{2mm}

In this paper, we complete this construction begun in \cite{Matsunaga:2013mba} by determining these additional terms which are necessitated for the nonlinear gauge invariance and by giving closed-form expressions for the action and nonlinear gauge transformations in the NS-NS sector of closed superstring field theory. 
We propose the action
\begin{align}
S= \int_{0}^{1}{dt} \, \langle \eta  \Psi _{t} ,  \, \mathcal{G}_{L} (t) \rangle  ,
\end{align}
where $\Psi _{t}$ is an NS-NS string field $\Psi $ plus $\tilde{\eta }$-exact terms and $\mathcal{G}_{L}$ is a pure-gauge solution to the NS {\it heterotic} string equation of motion in {\it the small Hilbert space of right movers}. 
The action has the WZW-like form and the almost same algebraic properties as the large-space action for NS open and NS closed (heterotic) string field theory \cite{Berkovits:2004xh}.

\vspace{2mm}

This paper is organized as follows. 
In section 2 we show that cubic and quartic actions can be determined by adding appropriate terms and imposing gauge invariance. 
In section 3, we briefly review the method of gauge-invariant insertions of picture-changing operators \cite{Erler:2013xta, Erler:2014eba} and provide some useful properties of the $(-,{\rm NS})$ closed superstring products. 
In section 4, first, we give the defining equation of $\mathcal{G}_{L}$ and associated fields which are necessitated to construct the NS-NS action. 
Then, we derive the WZW-like expression for the action and nonlinear gauge transformations and show that $\eta \, \mathcal{G}_{L} =0$ gives the NS-NS superstring equation of motion just as other large-space theories. 
We end with some conclusions.

\section{Nonlinear gauge invariance} %{Adding terms and Acquiring gauge invariance} %(3-8)

Let $\kappa $ be the coupling constant of closed string fields. 
We expand an action $S$ for NS-NS string field theory in powers of $\kappa $: $S=\frac{2}{\alpha ^{\prime }}  \sum_{n} \kappa ^{n} S_{n+2}$.       
In the large Hilbert space, which includes the $\xi$- and $\tilde{\xi}$-zero modes coming from bosonization of the $\beta \gamma$- and $\tilde{\beta } \tilde{\gamma}$-systems, the NS-NS string field $\Psi $ is a Grassmann even, (total) ghost number 0, left-mover picture number 0, and right-mover picture number 0 state.  
The free action $S_{2}$ is given by 
\begin{align}
S_{2} = \frac{1}{2} \langle \eta \Psi , Q \, \tilde{\eta } \Psi \rangle ,
\end{align}
where $Q$ is the BRST operator, $\eta$ is the zero mode of the left-moving current $\eta (z)$, and $\tilde{\eta }$ is the zero mode of the right-moving current $\tilde{\eta }(\tilde{z})$ \cite{Berkovits:1998bt}. 
The bilinear is the $c_{0}^{-}$-inserted BPZ inner product: $\langle A , B \rangle \equiv \langle \mathrm{bpz}(A) | c_{0}^{-} | B \rangle $, where $c_{0}^{-} = \frac{1}{2} ( c_{0} - \tilde{c}_{0})$. 
For brevity, we use the symbol $({\bf G} | {\bf p } , {\bf \tilde{p} } )$ which denotes that the total ghost number is ${\bf G}$, the left-mover picture number is ${\bf p}$, and the right-mover picture number is ${ \bf \tilde{p} }$. 
Then, ghost-and-picture numbers of $\Psi$, $Q$, $\eta $, and $\tilde{\eta }$ are $(0|0,0)$, $(1|0,0)$, $(1|-1,0)$, and $(1|0,-1)$ respectively.  
Note that the inner product $\langle A , B \rangle $ gives a nonzero value if and only if the sum of $A$'s and $B$'s ghost-and-picture numbers is equal to $( {\bf G} | {\bf p} , {\bf \tilde{p} } ) = (3 | -1 , -1)$. 
%%%%%%%%%%%%%%%%%%%%
Computing the variation of this action $\delta S_{2} = \langle \delta \Psi , \eta  Q \tilde{\eta } \Psi \rangle $, we obtain the equation of motion $Q \eta \tilde{\eta } \Psi =0$ and find that $S_{2}$ is invariant under the gauge transformation
\begin{align}
\delta _{1} \Psi = Q \, \Lambda + \eta \, \Omega + \tilde{\eta } \, \widetilde{\Omega },
\end{align}
where $\Lambda $, $\Omega$, and $\widetilde{\Omega }$ are gauge parameters of $Q$-, $\eta $, and $\tilde{\eta }$-gauge transformations respectively. 

\vspace{2mm}

We would like to construct nonzero and nonlinear gauge-invariant interacting terms $S_{3}$, $S_{4} $, $S_{5} $, $\dots $ using string fields $\Psi $ belonging to the large Hilbert space. 
In the kinetic term $S_{2}$, there exist three generators of gauge transformations: $Q$, $\eta$, and $\tilde{\eta }$. 
However, as we will see, only $Q$- and $\eta $-gauge invariances are extended to be nonlinear and $\tilde{\eta }$-gauge invariance remains to be linear in our interacting theory. 
Consequently, with two nonlinear and one trivial gauge invariances, the theory is Wess-Zumino-Witten-likely formulated. 

\vspace{2mm} 

In section 2, starting with the action proposed in \cite{Matsunaga:2013mba} and adding appropriate terms at each order of $\kappa$, we construct cubic and quartic terms $S_{3},S_{4}$ of the action, whose nonlinear gauge transformations of $Q$ and $\eta $ take WZW-like forms just as other large-space theories. 

\subsection{Cubic vertex} %(4-5)
\label{sec 2.1}

Let $\xi $ and $\tilde{\xi }$ be the zero modes of $\xi (z)$- and $\tilde{\xi } ( \tilde{z} )$-ghosts respectively, and $X$ and $\widetilde{X}$ be the zero modes of left- and right-moving picture-changing operators respectively. 
The $(n+2)$-point interaction term $S_{n+2}$ proposed in \cite{Matsunaga:2013mba} includes $\langle \eta \Psi , \big[   (Q \widetilde{Q} \Psi )^{n} ,  \tilde{\eta } \Psi \big] \rangle $ to correspond to the result of first quantization, where $[A_{1} , \dots , A_{n} ]$ is the bosonic string $n$-product and $\widetilde{Q} := \tilde{\eta } \tilde{\xi } \cdot Q \cdot \tilde{\xi } \tilde{\eta }$ is a projected BRST operator. 
This term becomes $\langle \xi \mathcal{V} , \big[  (X \widetilde{X} \mathcal{V} )^{n} , \tilde{\xi } \mathcal{V} \big] \rangle$ under partial gauge fixing $\Psi = \xi \tilde{\xi } \, \mathcal{V} _{ (2|-1,-1) }$. 
However, naive use of this term makes nonlinear gauge invariance not clear. 
%%%%%%%%%%%%%%%%%%%%
In this subsection, as the simplest example, we show that a gauge-invariant cubic term $S_{3}$ can be obtained by adding appropriate terms to
\begin{align}
\label{2.1-1}
\langle \eta \Psi , [ Q \widetilde{Q} \Psi , \tilde{\eta } \Psi ] \rangle  = \langle \eta \Psi ,  [ \widetilde{X} Q \, \tilde{\eta } \Psi , \tilde{\eta } \Psi ] \rangle .
\end{align}
%%%revision
%We write $\ld A , B \rd $ for the graded commutator $\ld A, B \rd \equiv A B - (-)^{AB} B A$, where the upper index of $(-)^{A}$ denotes the Grassmann parity of $A$. 
%Note that $X = \ld \xi , Q \rd $ and $\widetilde{X} = \ld \tilde{\xi } , Q \rd $. 
%\vspace{3mm}
%%%

\underline{Cyclicity}

\vspace{2mm}

It would be helpful to consider the cyclicity of vertices. 
A $(n+1)$-point vertex $V_{n} $ is called a (BPZ-) cyclic vertex if $V_{n}$ satisfies  
\begin{align}
\langle  A_{0} , V_{n} ( A_{1} , \dots , A_{n} )  \rangle   = (-) ^{ A_{0} ( A_{1} + \dots + A_{n} ) }  \langle  A_{1} , V_{n} ( A_{2} , \dots , A_{n} ,  A_{0} )  \rangle .
\end{align}
The upper index of $(-1)^{A}$ means the ghost number of $A$. 
When the $(n+1)$-point action $S_{n+1}$ is given by $S_{n+1} = \frac{1}{(n+1)!} \langle \Psi , V_{n}( \Psi ^{n} ) \rangle $ using a cyclic vertex $V_{n}$, its variation becomes 
\begin{align}
\delta S_{n+1} = \frac{1}{n!} \langle  \delta \Psi , V_{n}( \Psi ^{n} ) \rangle  .
\end{align}
Then, if there exist a zero divisor of the state $V_{n} ( \Psi ^{n} )$, it generates gauge transformations. 
For example, bosonic string field theories and superstring field theories in the small Hilbert space are the case and their gauge transformations are determined by $L_{\infty}$- or $A_{\infty}$-algebras. 
 
Next, we consider the following case: a vertex $V_{n}^{\prime }$ is not cyclic but has the following property
\begin{align}
\label{pseudo}
\delta S_{n+1} = \frac{1}{n!} \langle  \delta \Psi ,  V_{n}^{\prime } ( \Psi ^{n} ) + W_{n} (\Psi ^{n} ) \rangle .
\end{align}
Nonzero $W_{n}$ implies that the cyclicity of $V_{n}$ is broken. 
For instance, this relation holds when $V_{n}^{\prime }$ consists of a cyclic vertex with operator insertions: $V^{\prime }_{n}(\Psi ^{n}) = V_{n}( Q\Psi ^{k} , \Psi ^{n-k} )$ or $V_{n}^{\prime}$ consists of some combination of cyclic vertices $V^{\prime }_{n} ( \Psi ^{n} )  = V_{m+1} ( V_{n-m} ( \Psi ^{n-m} ) , \Psi ^{m} ) $. 
%%%%%%%%%%%%%%%%%%%%
In this case, in general, a zero divisor of $V_{n}^{\prime } + W_{n}$ gives the generator of gauge transformations. 
However, when $W_{n}$ consists of lower $V_{k<n}$, there exists a special case that the zero divisor of $V_{n}$ gives the generator of gauge transformations just as WZW-like formulations of superstring field theories in the large Hilbert space, which is the subject of this paper.

\vspace{3mm}

\underline{Adding terms}

\vspace{2mm}

We know that {\it naive} insertions of operators which do not work as derivations of string products, such as $\xi $, $\tilde{\xi }$, $X$, $\widetilde{X}$, and $\widetilde{Q}$, makes nonlinear gauge invariances not clear. 
We show that a cubic vertex satisfying the special case of $({\ref{pseudo}})$ can be constructed by adding appropriate terms and by imposing gauge invariances. 
Computing the variation of $(\ref{2.1-1})$, we obtain
\begin{align}
\delta \Big( \langle \eta \Psi , [ Q \widetilde{Q} \Psi , \tilde{\eta } \Psi ] \rangle \Big) 
& = \langle \delta \Psi , \eta [ Q \widetilde{Q} \Psi , \tilde{\eta } \Psi ] + \widetilde{Q} [ Q \Psi , \tilde{\eta } \Psi ] - [ Q \tilde{\eta } \Psi , \widetilde{Q} \Psi ] \rangle 
\\ \nonumber
& = \langle \delta \Psi , \eta \Big{\{ }   2 [ Q \widetilde{Q} \Psi , \tilde{\eta } \Psi ] + \widetilde{Q} [ Q \Psi , \tilde{\eta } \Psi ] \Big{\} }  +  \Big{\{ } 2 \widetilde{Q} [ \Psi , Q \tilde{\eta } \eta \Psi ] - [ \tilde{\eta } \Psi , \eta Q \widetilde{Q} \Psi   \Big{\} } \rangle .
\end{align}
We therefore add the term $\langle \eta \Psi , \widetilde{Q} [ Q \Psi , \tilde{\eta } \Psi ] \rangle $ for $(\ref{pseudo})$. 
The variation of this term becomes
\begin{align}
\delta \Big( \langle \eta \Psi ,  \widetilde{Q} [ Q \Psi , \tilde{\eta } \Psi ] \rangle \Big) 
&= \langle \delta \Psi , \eta \widetilde{Q} [ Q \Psi , \tilde{\eta } \Psi ] + Q [ \eta \widetilde{Q} \Psi , \tilde{\eta } \Psi ] - \tilde{\eta } [ \eta \widetilde{Q} \Psi  , Q \Psi ] \rangle 
\\ \nonumber 
&= \langle \delta \Psi , 
\eta \Big{\{ }  \widetilde{Q} [ Q \Psi , \tilde{\eta } \Psi ] + 2 [ Q \tilde{\eta } \Psi , \widetilde{Q} \Psi ]  \Big{\} }  
+  \Big{\{ } 2 [ Q \tilde{\eta } \eta \Psi , \widetilde{Q} \Psi ] - [ \tilde{\eta } \Psi , \eta Q \widetilde{Q} \Psi \Big{\} } \rangle .
\end{align}
Hence, the term $\langle \eta \Psi , [ Q \tilde{\eta } \Psi , \widetilde{Q} \Psi ] \rangle $ is necessitated for the property $(\ref{pseudo})$.  
The variation of this term is given by 
\begin{align}
\delta \Big(  \langle \eta \Psi , [ Q \tilde{\eta } \Psi , \widetilde{Q} \Psi ] \rangle \Big) 
&=  \langle \delta \Psi , \eta [ Q \tilde{\eta } \Psi ,\widetilde{Q} \Psi ] +  Q \tilde{\eta } [ \eta \Psi , \widetilde{Q} \Psi ] - \widetilde{Q}[ \eta \Psi , Q \tilde{\eta } \Psi ]  \rangle 
\\ \nonumber 
&= \langle \delta \Psi , \eta \Big{\{ }  [ Q \widetilde{Q} \Psi , \tilde{\eta } \Psi ]  +  \widetilde{Q} [ Q \Psi , \tilde{\eta } \Psi ] + [ Q \tilde{\eta } \Psi , \widetilde{Q} \Psi ]  \Big{\} }  
\\ \nonumber 
& \hspace{20mm} +  \Big{\{ }  \widetilde{Q} [ \Psi , Q \tilde{\eta } \eta \Psi ] - [ \widetilde{Q} \Psi , Q \tilde{\eta } \eta \Psi ] - [ \tilde{\eta } \Psi , \eta Q \widetilde{Q} \Psi  \Big{\} } \rangle .
\end{align}
Averaging these three terms, we obtain the cubic action satisfying $(\ref{pseudo})$
\begin{align}
\nonumber 
S_{3} =  \frac{ 1 }{3!} \langle \eta \Psi , \frac{1}{3} \Big( 
\widetilde{Q} [ Q \Psi , \tilde{\eta } \Psi ] + [ Q \widetilde{Q} \Psi , \tilde{\eta } \Psi ] + [Q \tilde{\eta } \Psi , \widetilde{Q} \Psi ] \Big) \rangle . 
%%%
%\\
%= & \frac{ 1 }{3!} \langle  \eta \Psi , \frac{1}{3} \Big(  \widetilde{X} [ Q \tilde{\eta } \Psi , \tilde{\eta } \Psi ] + [ \widetilde{X} Q \tilde{\eta } \Psi , \tilde{\eta } \Psi ] + [ Q \tilde{\eta } \Psi , \widetilde{X} \tilde{\eta } \Psi ]  \Big) \rangle  .
\end{align}
The variation of this cubic action is given by 
\begin{align}
\label{second line}
\nonumber 
\delta S_{3} & = \frac{ 1 }{2} \langle \delta \Psi , \frac{\eta }{3} \Big(  \widetilde{Q} [ Q \Psi , \tilde{\eta } \Psi ] + [ Q \widetilde{Q} \Psi , \tilde{\eta } \Psi ] + [ Q \tilde{\eta } \Psi , \widetilde{Q} \Psi ]  \Big) \rangle 
\\ \nonumber 
& \hspace{20mm} + \frac{ 1 }{2} \langle \delta \Psi , \frac{1}{3} \Big( \widetilde{Q} [ \Psi , \eta Q \tilde{\eta } \Psi ] - [ \widetilde{Q} \Psi , \eta Q \tilde{\eta } \Psi ] - [ \tilde{\eta } \Psi , \eta Q \widetilde{Q} \Psi ]  \Big) \rangle 
\\ 
& =   \frac{ 1 }{2} \langle \delta \Psi , V_{3} ( \Psi ^{3}) \rangle 
 - \frac{ 1 }{2} \langle \delta \Psi ,  \frac{1}{3} \Big( \widetilde{X} [ \tilde{\eta } \Psi , V_{1} ] + [ \widetilde{X} \tilde{\eta } \Psi , V_{1} ] + [ \tilde{\eta } \Psi , \widetilde{X} V_{1} ]  \Big) \rangle 
\end{align}
Note that $V_{1} = Q \tilde{\eta } \eta \Psi $ appears in $W_{2} \equiv \frac{1}{3} \Big( \widetilde{Q} [ \Psi , V_{1} ] - [ \widetilde{Q} \Psi , V_{1} ] - [ \tilde{\eta } \Psi , \widetilde{X} V_{1} ]  \Big) $ if one only if we use $ \widetilde{X} \tilde{\eta } = \ld Q, \tilde{\xi } \rd  \tilde{\eta }$ instead of $\widetilde{Q}$, where $\ld Q ,\xi \rd := Q \xi - (-)^{Q \xi } \xi Q$ is the graded commutator.

\vspace{3mm}

\underline{Gauge invariance $\delta _{2} S_{2} + \delta _{1} (\kappa S_{3})$}

\vspace{2mm}

Let us determine second order gauge transformation $\delta _{2} \Psi$ satisfying $\delta _{2} S_{2} + \delta _{1} (\kappa S_{3}) =0$.  
For this purpose, it is rather suitable to use a pair of fundamental operators $(Q, \eta, \tilde{\eta }, \tilde{\xi } )$ than to use a pair of composite operators such as $(Q , \eta , \tilde{\eta }, \widetilde{Q})$. 
For example, $V_{1}(\Psi )= \eta Q \tilde{\eta } \Psi $ appears in $W_{2} (\Psi ^{2})$ if and only if we use $(Q, \eta , \tilde{\eta } , \tilde{\xi } )$ and furthermore, while $\langle  \widetilde{Q} [ Q \Psi , \Lambda ] , \eta Q \tilde{\eta } \Psi \rangle = 0$, $\langle \widetilde{X} [ Q \tilde{\eta } \Psi , \Lambda ] , \eta Q \tilde{\eta } \Psi \rangle \not= 0 $. 
The first order $Q$-gauge transformation of $S_{3}$ is given by
\begin{align}
\delta _{1,\Lambda } S_{3} = \frac{ 1 }{2} \langle Q\Lambda , 
\frac{1}{3} \Big( \widetilde{X} [ Q \tilde{\eta } \Psi , \tilde{\eta } \eta \Psi ] + [ \widetilde{X} Q \tilde{\eta } \Psi , \tilde{\eta } \eta \Psi ] + [ Q \tilde{\eta } \Psi , \widetilde{X} \tilde{\eta } \eta \Psi ] \Big)  \rangle . 
\end{align}
Note that the zero mode $\widetilde{X}$ of the right-mover picture-changing operator is inserted cyclicly. 
We find that under the following second order gauge transformation
\begin{align}
\nonumber 
\delta _{2,\Lambda } \Psi = & \frac{\kappa }{3} \Big{\{ }  \widetilde{X} [ Q \tilde{\eta } \Psi , \Lambda ] + [ \widetilde{X} Q \tilde{\eta } \Psi , \Lambda ] + [ Q \tilde{\eta } \Psi ,  \widetilde{X} \Lambda ] \Big{\} } 
\\
& \hspace{20mm} - \frac{\kappa }{6} \Big{\{ } \widetilde{X} [ \tilde{\eta } \Psi , Q \Lambda ] +  [ \widetilde{X} \tilde{\eta } \Psi , Q \Lambda ] +  [ \tilde{\eta } \Psi , \widetilde{X} Q \Lambda ]     \Big{\} }  ,
\end{align}
the cubic term $S_{3}$ of the action is gauge invariant: $\delta _{2,\Lambda }S_{2} + \delta _{1, \Lambda } ( \kappa S_{3}) =0$.  
For brevity, we define the following new string product which includes the zero mode $\widetilde{X}$ of the right-mover picture-changing operator
\begin{align}
\label{two product}
[ A , B ]^{L} := \frac{1}{3} \Big( \widetilde{X} [ A, B ] + [ \widetilde{X} A , B ] + [ A , \widetilde{X} B ] \Big) .
\end{align}
This new product $[A,B]^{L} $ satisfies the same properties as original product $[A,B]$, namely, symmetric property $[A,B]^{L} = (-)^{AB} [ B ,A ] ^{L}$ and derivation properties of $Q$, $\eta $, and $\tilde{\eta }$. 
Note that when we use this new product, the cubic term $S_{3}$ of the action is given by
\begin{align}
S_{3}= \frac{1 }{3!} \langle \eta \Psi , [  Q \tilde{\eta } \Psi , \tilde{\eta } \Psi ]^{L} \rangle ,
\end{align} 
and the variation $\delta S_{3}$ becomes
\begin{align}
\label{delta S_{3}}
\delta S_{3} = \frac{1 }{2} \langle  \delta \Psi , \eta [ Q \tilde{\eta } \Psi , \tilde{\eta } \Psi ]^{L} +  [ \eta Q \tilde{\eta } \Psi , \tilde{\eta } \Psi ]^{L} \rangle .
\end{align}
Then, we can quickly show that $S_{2} + \kappa S_{3}$ is gauge invariant up to $O(\kappa ^{2})$ 
\begin{align}
\delta _{2, \Lambda } S_{2} + \delta _{1,\Lambda } ( \kappa S_{3} ) = \langle \delta _{2, \Lambda } \Psi , \eta Q \tilde{\eta } \Psi \rangle + \frac{\kappa }{2} \langle \delta _{1, \Lambda }\Psi  , [ Q \tilde{\eta } \Psi , \tilde{\eta } \eta \Psi ]^{L}  \rangle =0 ,
\end{align}
under the following $Q$-gauge transformations 
\begin{align}
\delta _{1, \Lambda } \Psi = Q \Lambda ,  \hspace{5mm}
\delta _{2, \Lambda } \Psi = \kappa [ Q \tilde{\eta } \Psi , \Lambda ]^{L} - \frac{\kappa }{2} [ \tilde{\eta } \Psi , Q \Lambda ]^{L} .
\end{align}

Similarly, we find that $S_{2}+ \kappa S_{3}$ is gauge invariant under $\eta$- and $\tilde{\eta }$-gauge transformations
\begin{align}
\delta _{1, \Omega } \Psi = \eta \, \Omega ,  \hspace{5mm}
\delta _{2, \Omega } \Psi =  -\frac{\kappa }{2} [ \tilde{\eta } \Psi , \eta \Omega ]^{L} , 
\\ 
\delta _{1, \widetilde{\Omega }} \Psi = \tilde{\eta} \, \widetilde{\Omega } ,   \hspace{5mm}
\delta _{2, \widetilde{\Omega }} \Psi = -\frac{\kappa }{2} [ \tilde{\eta } \Psi , \tilde{\eta } \widetilde{\Omega } ]^{L} , 
\end{align}
for example, as follows 
\begin{align}
\label{eta-gauge 2}
\delta _{2,\Omega } S_{2} + \delta _{1,\Omega } ( \kappa S_{3} ) = \langle \delta _{2, \Omega } \Psi , \eta Q \tilde{\eta } \Psi \rangle + \frac{\kappa }{2} \langle \delta _{1, \Omega } \Psi , [ Q \tilde{\eta } \Psi , \tilde{\eta } \eta \Psi ]^{L}  \rangle = 0 .
\end{align}

\vspace{2mm}

\underline{Linear gauge invariance} %$\delta _{\widetilde{\Omega }^{\prime }} S_{2} = \delta _{\widetilde{\Omega }^{\prime }} S_{3} = 0$}

\vspace{1mm}

In the same way as $(\ref{eta-gauge 2})$, the action $S_{2} + \kappa S_{3}$ is invariant under %$\tilde{\eta }$-gauge transformations 
\begin{align}
\delta _{1, \widetilde{\Omega }} \Psi = \tilde{\eta} \, \widetilde{\Omega } ,   \hspace{5mm}
\delta _{2, \widetilde{\Omega }} \Psi = -\frac{\kappa }{2} [ \tilde{\eta } \Psi , \tilde{\eta } \widetilde{\Omega } ]^{L} . 
\end{align}
Although it naively looks like nonlinear $\tilde{\eta }$-gauge invariance of the action, it is essentially {\it linear}. 
Note that the second order $\tilde{\eta }$-gauge transformation $\delta _{2, \widetilde{\Omega }} \Psi = \tilde{\eta } \big(- \frac{\kappa }{2} [ \tilde{\eta } \Psi , \widetilde{\Omega } ]^{L} \big) $ as well as the first order one $\delta _{1,\widetilde{\Omega }} \Psi = \tilde{\eta } \widetilde{\Omega }$ is $\tilde{\eta }$-exact. 
As a result, after the redefinition of the $\tilde{\eta }$-gauge parameter 
\begin{align}
\widetilde{\Omega }^{\prime } \equiv \widetilde{\Omega } - \frac{\kappa }{2} [ \tilde{\eta } \Psi , \widetilde{\Omega } ]^{L} + \dots  , 
\end{align}
the $\tilde{\eta }$-gauge transformation $\delta _{\widetilde{\Omega }^{\prime }} \Psi \equiv \delta _{1, \widetilde{\Omega }} \Psi + \delta _{2, \widetilde{\Omega }} \Psi + \dots $ becomes {\it linear} 
\begin{align}
\delta _{\widetilde{\Omega }^{\prime }} \Psi = \tilde{\eta } \, \widetilde{\Omega }^{\prime } . 
\end{align}
One can quickly check that $\delta _{\widetilde{\Omega }} S_{2} = \delta _{\widetilde{\Omega }} S_{3} = 0 $ holds, or equivalently, the second input of $(\ref{delta S_{3}})$ is an $\tilde{\eta }$-exact state. 
Hence, $S_{2} + \kappa S_{3}$ still remains invariant under the {\it linear} $\tilde{\eta }$-gauge transformation.

\subsection{Quartic vertex} %(6-8)

We can construct the quartic term $S_{4}$ and, in principle, higher interaction terms $S_{n>4}$ by repeating the same procedure. 
More precisely, starting with $\langle \eta \Psi , [ ( \widetilde{X} Q \tilde{\eta } \Psi )^{2}, \tilde{\eta } \Psi ] \rangle$, adding appropriate terms for $(\ref{pseudo})$, and imposing the gauge invariance $\delta _{3} S_{2} + \delta _{2} ( \kappa  S_{3} ) + \delta _{1} ( \kappa ^{2} S_{4} )=0$, we obtain the quartic term $S_{4}$. 
First, we consider the gauge invariance $\delta _{3} S_{2} + \delta _{2} ( \kappa  S_{3} ) + \delta _{1} ( \kappa ^{2} S_{4} )=0$.

\vspace{3mm}

\underline{To be $\delta _{3} S_{2} + \delta _{2} ( \kappa  S_{3} ) + \delta _{1} ( \kappa ^{2} S_{4} )=0$ } 

\vspace{2mm}

To be gauge invariant, the first order gauge transformation of $\kappa ^{2} S_{4}$ has to cancel $\delta _{3} S_{2} + \delta _{2} (\kappa S_{3})$. 
Note that $\delta _{2} (\kappa S_{3})$ is given by 
\begin{align}
\label{2.2-1}
\delta _{2} (\kappa S_{3}  ) = \frac{\kappa ^{2}}{4}  \langle \Lambda , 2 \big[  [ Q \tilde{\eta } \Psi , \tilde{\eta } \eta \Psi ]^{L} , Q \tilde{\eta } \Psi \big] ^{L} 
+  Q \big[  [ Q \tilde{\eta } \Psi ,  \tilde{\eta } \eta \Psi ]^{L} , \tilde{\eta } \Psi  \big] ^{L}  \rangle .
%+ [ [ Q\eta \Psi , Q \eta \bar{\eta } \Psi ]^{L} \eta \Psi ]^{L} + [ [ Q \eta \Psi , \eta \bar{\eta } \Psi ]^{L} , Q \eta \Psi ]^{L} \rangle 
\end{align}
Therefore, we have to consider the following terms 
\begin{align}
& \big[    [  A ,  B ]^{L}    ,  C    \big] ^{L} = \frac{1}{3 \cdot 3} \Big(  \widetilde{X} \big[  [ \widetilde{X} A ,  B ] + [  A , \widetilde{X} B ] , C \big] +\big[ [ \widetilde{X} A ,  B ] + [ A , \widetilde{X} B ] , \widetilde{X} C \big]   \Big) 
\\ \nonumber 
& \hspace{8mm} + \frac{1}{3 \cdot 3} \Big(   \big[ \widetilde{X} [  \widetilde{X} A ,  B ] + \widetilde{X} [  A ,  \widetilde{X} B ] , C \big] +\big[  \widetilde{X} [  A ,  B ] , \widetilde{X} C \big] + \widetilde{X} \big[  \widetilde{X} [ A ,B ] , C \big]  + \big[ \widetilde{X}^{2} [  A ,  B ] , C \big] \Big) , 
\end{align}
where $A,B,C=$ $\tilde{\eta } \Psi $, $Q \tilde{\eta } \Psi $, $\eta \tilde{\eta } \Psi $, or $Q \eta \tilde{\eta } \Psi$. 
%%%%%%%%%%%%%%%%%%%%
To cancel the second line by acting $Q$, the quartic term $S_{4}$ has to include the following types of terms 
\begin{align}
\nonumber 
& \tilde{\xi } \big[  \widetilde{X} [ A, B ], C \big] ,  \hspace{5mm} \big[ \widetilde{X} [ \tilde{\xi } A ,B ] , C \big] , \hspace{5mm} \big[ \widetilde{X} [ A, \tilde{\xi } B ] , C \big] , \hspace{5mm} \big[ \widetilde{X} [ A , B ] , \tilde{\xi } C \big] , \hspace{5mm} \big[ \tilde{\xi } \widetilde{X} [ A, B ] , C \big] , 
\\ \nonumber 
& \hspace{18mm}  \widetilde{X} \big[ \tilde{\xi } [ A , B ] , C \big] ,  \hspace{5mm} \big[ \tilde{\xi } [ \widetilde{X} A , B ] , C \big] ,  \hspace{5mm} \big[ \tilde{\xi } [ A , \widetilde{X} B ] , C \big] , \hspace{5mm} \big[ \tilde{\xi } [ A , B ] , \widetilde{X} C \big] .
\end{align}
Of course, we can repeat similar computations of above terms as we did in subsection $\ref{sec 2.1}$. 
However, there exists a reasonable shortcut.  
%%%%%%%%%%%%%%%%%%%%
Note that, for example, the following relation holds
\begin{align}
\nonumber 
&Q \Big( \widetilde{X} \big[  \tilde{\xi } [ A ,B ] , C \big]   \Big) +  \widetilde{X}  \big[ \tilde{\xi } [ Q A  , B ] , C \big]  + (-)^{A} \widetilde{X} \big[ \tilde{\xi } [ A , Q B ] , C \big]  
\\ \nonumber 
& \hspace{35mm} + (-)^{A+B} \widetilde{X} \big[ \tilde{\xi } [ A , B ] , Q C  \big]  +  \Big( \widetilde{X} \big[ \widetilde{X} [A, B ] , C \big]  \Big) =0 . 
\end{align}
Hence, the three product $L_{2+2}(A,B,C)$ defined by 
%%%%%%%%%%%%%%%%%%%%
%\begin{align}
%\label{2.2-2+2}
%\nonumber 
%L_{2+2} ( A, B, C ) := \, & \frac{1}{9 \cdot 2} \Big{\{ }   X \big[ \xi [ A,B ] , C \big] + \big[ \xi \big( [ X A , B ] + [ A ,X B ] \big)  , C \big] + \big[ \xi [ A ,B ] , X C \big] 
%\\ \nonumber 
%& \hspace{11mm} - \xi \big[ X [ A,B ] , C \big] - \big[ X \big( [ \xi A , B ] + (-)^{A} [ A , \xi B ] \big) , C \big] 
%\\ \nonumber 
%& \hspace{14mm} - (-)^{A+B} \big[ X [ A ,B ] , \xi C \big]  + 2 \big[ \xi X [ A ,B ] , C \big]  
%\\ \nonumber 
%& \hspace{11mm} - X \big[ [ \xi A , B ] + (-)^{A} [ A , \xi B ] , C \big] - \big[ [ \xi A, B ] + (-)^{A} [ A , \xi B ] , X C \big]  
%\\ \nonumber 
%& \hspace{11mm} - \xi \big[ [ X A , B ] + [ A , X B ] , C \big] - (-)^{A+B} \big[ [ X A , B ] + [ A , X B ] , \xi C \big]  \Big{\} } 
%\\ 
%& \hspace{5mm} + \big{\{ } ( B, C,A )  { \mathchar`- }  \mathrm{terms} \big{\} }+ \big{\{ } (C,A,B ) { \mathchar`-}  \mathrm{terms} \big{\} } 
%\end{align}
\begin{align}
\label{2.2-2+2}
\nonumber 
L_{2 + 2} ( A , B , C ) := \, & \frac{1}{9 \cdot 2} \Big{\{ } 2 \big[ \tilde{\xi } \widetilde{X} [ A ,B ] , C \big]  - \tilde{\xi } \big[ \widetilde{X} [ A,B ] + [ \widetilde{X} A , B ] + [ A , \widetilde{X} B ] , C \big]  
\\ \nonumber 
& \hspace{9mm} + \widetilde{X} \big[ \tilde{\xi } [ A , B ] , C \big] + \big[ \tilde{\xi } \big( [ \widetilde{X} A , B ] + [ A , \widetilde{X} B ] \big)  , C \big] + \big[ \tilde{\xi } [ A ,B ] , \widetilde{X} C \big] 
\\ \nonumber 
& \hspace{11mm} - \widetilde{X} \big[ [ \tilde{\xi } A , B ] , C \big]  - \big[ \widetilde{X}  [ \tilde{\xi } A , B ] , C \big]  - \big[ [ \tilde{\xi } A , B ] , \widetilde{X} C \big]  
\\ \nonumber 
& \hspace{9mm} - (-)^{A} \Big( \big[  \widetilde{X} [ A , \tilde{\xi } B ] , C \big] + \widetilde{X} \big[ [ A , \tilde{\xi } B ] , C \big] + \big[ [ A , \tilde{\xi } B ] , \widetilde{X} C \big] \Big) 
\\ \nonumber 
& \hspace{11mm} - (-)^{A + B} \Big( \big[ \widetilde{X} [ A ,B ]  + [ \widetilde{X} A , B ] + [ A , \widetilde{X} B ] , \tilde{\xi } C \big] \Big) \Big{\} } 
\\ 
& \hspace{5mm} + \big{\{ } ( B , C , A )  { \mathchar`- }  \mathrm{terms} \big{\} }+ \big{\{ } ( C , A , B ) { \mathchar`-}  \mathrm{terms} \big{\} } 
\end{align}
%%%%%%%%%%%%%%%%%%%%%%%
and the two product $[ A , B ]^{L}$ defined by $(\ref{two product})$ satisfy an $L_{\infty }$-relation up to $\mathcal{O}( \Psi ^{4} )$:
\begin{align}
\nonumber 
\label{2.3-1}
& Q L_{2+2} ( A , B, C) + L_{2+2} (QA,B ,C) +(-)^{A} L_{2+2} (A,QB,C) + (-)^{A+B} L_{2+2} (A,B,QC) 
\\
& \hspace{10mm} + \big[ [ A , B ]^{L} , C \big] ^{L} +(-)^{A(B+C)} \big[ [ B , C ]^{L} , A \big] ^{L} +(-)^{C(A+B)} \big[ [ C , A ]^{L} , B \big] ^{L} =0  . 
\end{align}
This new three product $L_{2+2} (A,B,C)$ possesses the symmetric property and the derivation property of $Q$. 
Note that, however, the operator $\eta $ does not act as a derivation of $L_{2+2} (A,B,C)$. 
To see this fact, we introduce the derivation-testing operation $\Delta _{\mathbb{X}} $ for $\mathbb{X} = Q, \eta , \tilde{\eta }$ 
\begin{align}
\nonumber 
\Delta _{\mathbb{X} } L_{2+2} (A,B, C) &:= \mathbb{X} L_{2+2} (A,B,C) + (-)^{\mathbb{X}}L_{2+2} ( \mathbb{X} A ,B ,C ) 
\\
& \hspace{10mm} + (-)^{\mathbb{X}(1+A)} L_{2+2} ( A ,\mathbb{X} B , C ) + (-)^{\mathbb{X} (1+A+B)}  L_{2+2} ( A, B ,\mathbb{X} C ) .
\end{align}
For example, $\Delta _{\mathbb{X} } [A,B ]^{L} = \mathbb{X} [A , B ] ^{L} +(-)^{\mathbb{X}} [ \mathbb{X} A ,B ] ^{L} +(-)^{\mathbb{X}(1+A)} [ A, \mathbb{X} B ]^{L} = 0$ holds for $\mathbb{X} = Q$, $\eta $, and $\tilde{\eta }$. 
Computing $\Delta _{\eta / \tilde{\eta }} L_{2+2} (A,B,C)$, we find $\Delta _{\eta } L_{2+2} ( A , B , C ) = 0$ but 
\begin{align}
\label{2.2-2}
\nonumber 
\Delta _{\tilde{\eta } } L_{2+2} ( A, B, C ) = \, & \frac{1}{3 \cdot 2} \Big{\{ }  \widetilde{X} \big[ [ A , B ] , C \big] + \big[ \tilde{\xi } \big( [ \widetilde{X} A , B ] + [ A , \widetilde{X} B ] \big)  , C \big] + \big[  [ A , B ] , \widetilde{X} C \big]  \Big{\} }
\\ 
& \hspace{5mm} - \frac{1}{3} \big[ \widetilde{X} [ A , B ] , C \big]   
 + \big{\{ } ( B , C , A )  { \mathchar`- }  \mathrm{terms} \big{\} }+ \big{\{ } (C,A,B ) { \mathchar`-}  \mathrm{terms} \big{\} } . 
\end{align}
$\Delta _{\tilde{\eta }} L_{2 + 2}( A , B , C ) \not= 0$ means that the $\tilde{\eta }$-derivation property is broken. 
However, by adding the appropriate term $L_{1+3}$ satisfying $\Delta _{Q/\eta } L_{1+3} (A,B,C)=0$, we can construct the three product satisfying the $L_{\infty }$-relation $(\ref{2.3-1})$ and $\eta$- and $\tilde{\eta }$-derivation properties $\Delta _{\eta / \tilde{\eta }} (L_{2+2} + L_{1+3})=0$.   

\vspace{2mm}

Note that the following types of products have the $Q$-derivation property
\begin{align}
\nonumber 
M(A,B,C) &= \big[ \widetilde{X} A , \widetilde{X} B , C \big] + \big[ [ \tilde{\xi } A , \widetilde{X} B ] , C \big] + (-)^{(A+1)(B+C)} \big[ [ \widetilde{X} B , C] , \tilde{\xi } A \big] 
\\ \nonumber 
& \hspace{30mm} + (-)^{C(A+B+1)} \big[ [ C , \tilde{\xi } A ] , \widetilde{X} B \big] ,
\\ \nonumber 
N(A,B,C) &= \widetilde{X} \big[ \tilde{\xi } [ A , B ] , C \big] + \tilde{\xi } \big[ \widetilde{X} [ A , B ] , C \big] ,
\end{align}
namely, $\Delta _{Q} M  = \Delta _{Q} N =0$. 
Therefore, $L_{1+3}$ is given by a linear combination of these $M$- and $N$-types of products, 
whose coefficients are fixed by the cancellation of the second line of $(\ref{2.2-2})$ and the sum of $N$-type products: 
\begin{align}
\label{2.2-1+3}
\nonumber 
L_{1+3} ( A, B, C ) := \, & \frac{1}{8\cdot 2} \Big{\{ } \widetilde{X}^{2} \big[ A , B , C \big] + \big[ \widetilde{X}^{2} A , B ,C \big] + \big[ A , \widetilde{X}^{2} B , C \big] + \big[ A , B , \widetilde{X}^{2} C \big]  \Big{\} } 
\\ \nonumber 
& \hspace{5mm}  + \frac{1}{8 \cdot 2} \Big{\{ } \tilde{\xi } \widetilde{X} \big[ [  A , B ] , C \big]  + \big[ [ \tilde{\xi } \widetilde{X} A ,  B ] , C \big] 
\\ \nonumber 
& \hspace{25mm} + (-)^{A} \big[ [ A, \tilde{\xi } \widetilde{X} B ] ,  C \big] + (-)^{A+B} \big[ [ A , B ] , \tilde{\xi } \widetilde{X} C \big] \Big{\} } 
\\ \nonumber 
& \hspace{5mm} + \frac{1}{8} \Big{\{ }  \widetilde{X} \big[ \widetilde{X} A , B , C \big] + \widetilde{X} \big[ A , \widetilde{X} B , C \big] + \widetilde{X} \big[ A , B , \widetilde{X} C \big] 
\\ \nonumber 
& \hspace{25mm} + \big[ \widetilde{X} A , \widetilde{X} B , C \big] + \big[ \widetilde{X} A , B , \widetilde{X} C \big] + \big[ A , \widetilde{X} B , \widetilde{X} C \big]  \Big{\} }  
\\ \nonumber 
& \hspace{5mm} - \frac{1}{8\cdot 2} \Big{\{ }  \tilde{\xi } \big[ [ \widetilde{X} A , B ] + [ A, \widetilde{X} B ] , C \big] + \tilde{\xi } \big[ [ A , B ] , \widetilde{X} C \big] 
\\ \nonumber 
& \hspace{20mm} +  \widetilde{X} \big[ [ \tilde{\xi } A , B ] , C \big]  -  \big[ [ \tilde{\xi } A , \widetilde{X} B ] , C \big]  - \big[ [ \tilde{\xi } A , B ] , \widetilde{X} C \big]  
\\ \nonumber 
& \hspace{20mm}  + (-)^{A} \Big( \widetilde{X} \big[ [ A, \tilde{\xi } B ] , C \big] - \big[ [ \widetilde{X} A , \tilde{\xi } B ] , C \big] - \big[ [ A , \tilde{\xi } B ] , \widetilde{X} C \big] \Big) 
\\ \nonumber 
& \hspace{20mm} + (-)^{A+B}  \Big( \widetilde{X} \big[ [ A, B ] , \tilde{\xi } C \big]  - \big[ [ \widetilde{X} A , B ] , \tilde{\xi } C \big] - \big[ [ A , \widetilde{X} B ] , \tilde{\xi } C \big] \Big)  \Big{\} } 
\\ \nonumber 
& \hspace{0mm} + \frac{1}{4\cdot 3} \Big{\{ }  \widetilde{X} \big[ \tilde{\xi } [ A , B ] , C \big] + [ \tilde{\xi } [ \widetilde{X} A , B ] + \tilde{\xi } [ A , \widetilde{X} B ] , C \big] + \big[ \tilde{\xi } [ A , B ] , \widetilde{X} C \big]  
\\ \nonumber 
& \hspace{1mm} + \tilde{\xi } \big[ \widetilde{X} [ A , B ] , C \big] + \big[ \widetilde{X} \big( [ \tilde{\xi } A , B ] + (-)^{A} [ A , \tilde{\xi } B ] \big) , C \big] + (-)^{A+B} \big[ \widetilde{X} [ A , B ] , \tilde{\xi } C \big] \Big{\} } 
\\ 
& \hspace{15mm} + \big{\{ }  ( B, C ,A ) {\mathchar`-} \mathrm{terms}  \big{\} } + \big{\{ }  (  C ,A , B ) \mathchar`- \mathrm{terms} \big{\} } .
\end{align}
$L_{1+3}(A,B,C)$ satisfies $\Delta _{Q} L_{1+3} (A,B,C) = 0$ because of $\Delta _{Q} M  = \Delta _{Q} N =0$. 
Hence, we define the following new three string product 
\begin{align}
[A,B,C]^{L} := L_{2+2}(A,B,C) + L_{1+3} (A,B,C) ,
\end{align} 
which satisfies the same properties as the original three product $[A,B,C]$, namely, the $L_{\infty }$-relation and the $\eta $- and $\tilde{\eta }$-derivation properties 
\begin{align}
\label{Q+J=0 3}
&\Delta _{Q} [ A ,B ,C ]^{L} + \big[ [ A , B ]^{L} , C \big] ^{L} + (-)^{A(B+C)} \big[ [ B , C ]^{L} , A \big] ^{L} + (-)^{C(A+B)} \big[ [ C ,A ]^{L} ,B \big] ^{L} = 0 ,
\\ 
& \hspace{45mm} \Delta _{\eta } [ A , B ,C ]^{L} = \Delta _{\tilde{\eta }} [ A , B , C ]^{L} = 0 .
\end{align}

\vspace{2mm}

Note also that the new product $[ A , B ,C ]^{L}$ includes %the following term 
%\begin{align}
$\langle \eta \Psi , [ \widetilde{X} Q \tilde{\eta } \Psi , \widetilde{X} Q \tilde{\eta } \Psi , \tilde{\eta } \Psi ] \rangle $ 
%\in \langle \eta \Psi , L_{1+3}( Q \tilde{\eta } \Psi , Q \tilde{\eta } \Psi , \tilde{\eta } \Psi ) \rangle . 
%\end{align} 
( for $A,B=Q \tilde{\eta } \Psi $, $C = \tilde{\eta } \Psi $) and 
this term becomes $\langle  \xi \mathcal{V} , [ X \widetilde{X} \mathcal{V} , X \widetilde{X} \mathcal{V} , \tilde{\xi } \mathcal{V} ]  \rangle$ under the partial gauge fixing $\Psi = \xi \tilde{\xi } \mathcal{V}$, which is necessitated for the correspondence to the result of first quantization.

\vspace{3mm}

\underline{Quartic vertex $S_{4}$}

\vspace{2mm}

Let us now consider the quartic vertex having the property $(\ref{pseudo})$ and determine the third order gauge transformation $\delta _{3} \Psi $. 
To obtain the gauge invariance $\delta _{3} S_{2} + \delta _{2} ( \kappa  S_{3} ) + \delta _{1} ( \kappa ^{2} S_{4} )=0$, the quartic term $S_{4}$ has to include the term $\langle \eta \Psi , [  Q \tilde{\eta } \Psi , Q \tilde{\eta } \Psi , \tilde{\eta } \Psi ]^{L} \rangle $ because the $L_{\infty }$-relation $(\ref{Q+J=0 3})$ is the only way to eliminate the term $\big[ [A,B]^{L} , C \big] ^{L}$ appearing in $\delta _{3} S_{2} + \delta _{2} ( \kappa  S_{3} )$. 
We therefore start with the following computation 
\begin{align}
\nonumber 
\delta \Big( \langle   \eta \Psi ,  [ Q \tilde{\eta } \Psi , Q \tilde{\eta } \Psi , \tilde{\eta } \Psi ]^{L}  \rangle \Big) = &
4 \langle \delta \Psi , \eta \Big{\{ } [ Q \tilde{\eta } \Psi , Q \tilde{\eta } \Psi , \tilde{\eta } \Psi ]^{L}  + \frac{1}{2} \big[  [ Q \tilde{\eta } \Psi , \tilde{\eta } \Psi ]^{L} , \tilde{\eta } \Psi \big] ^{L} \Big{\} } \rangle  
\\ \nonumber 
& + 4 \langle \delta \Psi,   2  [ \eta Q \tilde{\eta } \Psi ,  Q \tilde{\eta } \Psi , \tilde{\eta } \Psi  ]^{L} 
- \frac{1}{2} \big[   [ \eta Q \tilde{\eta } \Psi , \tilde{\eta } \Psi ]^{L}  ,  \tilde{\eta } \Psi  \big] ^{L}  \rangle 
\\
& \hspace{10mm} + 4 \langle \delta \Psi ,   \big[  [ \tilde{\eta } \eta \Psi ,  Q \tilde{\eta } \Psi  ]^{L} , \tilde{\eta } \Psi  \big] ^{L} \rangle .
\end{align}
To obtain the quartic vertex having the property $(\ref{pseudo})$, the term $\langle  \Psi , \eta  \big[  [ Q \tilde{\eta } \Psi , \tilde{\eta } \Psi ]^{L} , \tilde{\eta } \Psi \big] ^{L} \rangle$ is necessitated. 
The variation of this term becomes
\begin{align}
\delta \Big( \langle  \Psi , \eta  \big[  [ Q \tilde{\eta } \Psi , \tilde{\eta } \Psi ]^{L} , \tilde{\eta } \Psi \big] ^{L} \rangle  \Big) =
2 \langle \delta \Psi , \eta  \big[  [ Q \tilde{\eta } \Psi , \tilde{\eta } \Psi ]^{L} , \tilde{\eta } \Psi \big] ^{L}  +  \big[  [ \tilde{\eta } \eta \Psi ,  Q \tilde{\eta } \Psi  ]^{L} , \tilde{\eta } \Psi  \big] ^{L} \rangle .
\end{align}
Hence, the quartic term $S_{4}$ defined by
\begin{align}
S_{4} = \frac{1}{4!} \langle \eta \Psi , [ Q \tilde{\eta } \Psi , Q \tilde{\eta } \Psi , \tilde{\eta } \Psi  ]^{L}  + \big[  [ Q \tilde{\eta } \Psi , \tilde{\eta } \Psi ]^{L} , \tilde{\eta } \Psi \big] ^{L}  \rangle ,
\end{align}
has the property $(\ref{pseudo})$ and its variation is given by 
\begin{align}
\label{2.3-2}
\nonumber 
& \delta S_{4} =  \frac{1}{3!} \langle \delta \Psi , \eta \Big( \big[ Q \tilde{\eta } \Psi , Q \tilde{\eta } \Psi , \tilde{\eta } \Psi \big] ^{L} + \big[ [ Q \tilde{\eta } \Psi , \tilde{\eta } \Psi ]^{L} , \tilde{\eta } \Psi \big] ^{L} \Big)  \rangle 
+ \frac{1}{4} \langle \delta \Psi , \big[  [ Q \tilde{\eta } \Psi , \tilde{\eta } \eta \Psi ]^{L} , \tilde{\eta } \Psi  \big] ^{L} \rangle 
\\
& \hspace{20mm}  +  \frac{2}{4!}  \langle \delta \Psi , \big[  \tilde{\eta } \Psi , [ \tilde{\eta } \Psi , \eta Q \tilde{\eta } \Psi ]^{L}  \big] ^{L} \rangle 
+ \frac{1}{3} \langle \delta \Psi , \big[ Q \tilde{\eta } \Psi , \tilde{\eta } \Psi , \eta Q \tilde{\eta } \Psi \big] ^{L}  \rangle . 
\end{align}
Note that $W_{3} = \frac{3}{2}[ V_{2} , \tilde{\eta } \Psi  ]^{L} - [ W_{2} , \tilde{\eta } \Psi ]^{L} - 2 [ Q \tilde{\eta } \Psi , V_{1} , \tilde{\eta } \Psi  ] ^{L}$ (the second term and the second line), where  $V_{1} = \eta Q \tilde{\eta } \Psi $, $V_{2}  = \eta [ Q \tilde{\eta } \Psi , \tilde{\eta } \Psi ] ^{L}$ and $W_{2} = [ V_{1} , \tilde{\eta } \Psi  ] ^{L}$. 
Using this $\delta S_{4}$, we can determine the third order gauge transformation $\delta _{3} \Psi $ satisfying $\delta _{3} S_{2} + \delta _{2} ( \kappa  S_{3} ) + \delta _{1} ( \kappa ^{2} S_{4} )=0 $. 
Since the $Q$-gauge transformation $\delta _{2, \Lambda } ( \kappa S_{3} ) + \delta _{1,\Lambda }(\kappa ^{2} S_{4})$ is given by
\begin{align}
\nonumber 
& \frac{\kappa ^{2} }{2} \langle  \Lambda , \big[ Q \tilde{\eta } \Psi ,  Q \tilde{\eta } \Psi , \eta Q \tilde{\eta } \Psi \big] ^{L} \rangle  
+ \frac{\kappa ^{2}}{2} \langle \Lambda , \big[  [ Q \tilde{\eta } \Psi , \tilde{\eta } \Psi ]^{L} ,  \eta Q \tilde{\eta } \Psi \big] ^{L} \rangle 
+  \frac{\kappa ^{2}}{2} \langle  \Lambda ,  \big[ Q \tilde{\eta } \Psi ,  [ \tilde{\eta } \Psi , \eta Q \tilde{\eta } \Psi ]^{L}  \big] ^{L} \rangle 
\\
& \hspace{20mm}  +  \frac{\kappa ^{2} }{12}  \langle Q \Lambda , \big[  \tilde{\eta } \Psi , [ \tilde{\eta } \Psi ,  \eta Q \tilde{\eta } \Psi ]^{L}  \big] ^{L} \rangle 
+ \frac{\kappa ^{2} }{3} \langle Q \Lambda , \big[ \tilde{\eta } \Psi , Q \tilde{\eta } \Psi , \eta Q \tilde{\eta } \Psi \big] ^{L}  \rangle ,
\end{align}
we obtain the third order $Q$-gauge transformation
\begin{align}
\nonumber 
& \delta _{3, \Lambda } \Psi = \frac{\kappa ^{2} }{2}  [ Q \tilde{\eta } \Psi , Q \tilde{\eta } \Psi , \Lambda ]^{L} + \frac{\kappa ^{2}}{2} \big[ [ Q \tilde{\eta } \Psi , \tilde{\eta } \Psi ]^{L} , \Lambda \big] ^{L}  
- \frac{\kappa ^{2} }{2} \big[ \tilde{\eta } \Psi , [ Q \tilde{\eta } \Psi , \Lambda ]^{L} \big] ^{L}
\\ 
& \hspace{25mm} 
- \frac{\kappa ^{2}}{12} \big[ \tilde{\eta } \Psi , [ \tilde{\eta } \Psi , Q \Lambda ]^{L} \big] ^{L}  
- \frac{\kappa ^{2} }{3} [ \tilde{\eta } \Psi , Q \tilde{\eta } \Psi , Q \Lambda ]^{L} . 
\end{align}
Similarly, the third order $\eta $- and $\tilde{\eta }$-gauge transformations are given by 
\begin{align}
\delta _{3, \Omega } \Psi &=  -\frac{\kappa ^{2} }{12} \big[ \tilde{\eta } \Psi , [ \tilde{\eta } \Psi , \eta \Omega ]^{L} \big] ^{L} - \frac{\kappa ^{2} }{3} \big[ \eta \Omega , Q \tilde{\eta } \Psi , \tilde{\eta } \Psi  \big] ^{L} , 
\\
\delta _{3, \widetilde{\Omega }} \Psi &=  -\frac{\kappa ^{2} }{12} \big[ \tilde{\eta } \Psi , [ \tilde{\eta } \Psi , \tilde{\eta } \widetilde{\Omega } ]^{L} \big] ^{L} - \frac{\kappa ^{2} }{3} \big[ \tilde{\eta } \widetilde{\Omega } , Q \tilde{\eta } \Psi , \tilde{\eta } \Psi  \big] ^{L} .
\end{align}
Note, however, that since $\delta _{3, \widetilde{\Omega } } \Psi $ as well as $\delta _{2, \widetilde{\Omega } }\Psi $ is $\tilde{\eta }$-exact, redefining $\tilde{\eta }$-gauge parameters as
\begin{align}
\widetilde{\Omega }^{\prime } \equiv \widetilde{\Omega } - \frac{\kappa }{2} [ \tilde{\eta } \Psi , \widetilde{\Omega } ]^{L} - \frac{\kappa ^{2} }{3} \Big(  \frac{1}{4} \big[ \tilde{\eta } \Psi , [ \tilde{\eta } \Psi , \widetilde{\Omega } ]^{L} \big] ^{L} + \big[ Q \tilde{\eta } \Psi , \tilde{\eta } \Psi , \widetilde{\Omega } \big] ^{L} \Big) + \dots , 
\end{align}
the $\tilde{\eta }$-gauge transformation $\delta _{\widetilde{\Omega }^{\prime }} \Psi \equiv \delta _{1, \widetilde{\Omega }} \Psi + \delta _{2, \widetilde{\Omega }} \Psi + \delta _{3, \widetilde{\Omega }} \Psi +\dots $ becomes linear $\delta _{\widetilde{\Omega }^{\prime } }\Psi = \tilde{\eta } \widetilde{\Omega }^{\prime }$.

\vspace{3mm}

In principle, we can construct higher vertices $S_{5}, S_{6}, \dots $ by repeating these steps at each order of $\kappa $. 
However, it is not easy to read a closed form expression by hand calculation because higher order vertices consist of a lot of terms and each term has complicated operator insertions. 
To obtain a closed form expression of all vertices in an elegant way, we necessitate another point of view, which we explain in section 3.  

%\begin{align}
%\delta S_{4} = \, & \frac{1}{3!} \langle \delta \Psi , \bar{\eta } [ Q\eta \Psi , Q \eta \Psi , \eta \Psi ]^{L} + \bar{\eta } [ [Q\eta \Psi , \eta \Psi ]^{L} , \eta \Psi ]^{L} 
%\\
%& + [ \eta \Psi , [ \eta \Psi ,  \bar{\eta } Q \eta \Psi ]^{L}  ]^{L} + [  Q\eta \Psi , \eta \Psi , \bar{\eta } Q \eta \Psi ]^{L}
%\rangle 
%\end{align}
%\begin{align}
%\delta \Big( \langle  \Psi ,  \bar{\eta }   \big[  [ Q\eta \Psi , \eta \Psi ]^{L} , \eta \Psi \big] ^{L} \rangle  \Big) =
%2 \langle \delta \Psi ,   \bar{\eta }   \big[  [ Q\eta \Psi , \eta \Psi ]^{L} , \eta \Psi \big] ^{L}  +  \big[  [ \eta \bar{\eta } \Psi ,  Q\eta \Psi  ]^{L} , \eta \Psi  \big] ^{L} \rangle 
%\end{align}
%\begin{align}
%\delta \Big( \langle  \Psi , \bar{\eta } [  Q\eta \Psi , Q \eta \Psi , \eta \Psi ]^{L}  \Big) = &
%\langle \delta \Psi , 4  \bar{ \eta } [  Q\eta \Psi , Q \eta \Psi , \eta \Psi ]^{L}    + 2 \bar{ \eta } \big[  [ Q\eta \Psi , \eta \Psi ]^{L} , \eta \Psi \big] ^{L}  \rangle  
%\\
%& +  \langle \delta \Psi,   8 [  \bar{\eta } Q \eta \Psi ,  Q\eta \Psi , \eta \Psi   ]^{L} -% 2 \big[   [  \bar{\eta } Q \eta \Psi , \eta  \Psi ]^{L}  ,   \eta \Psi  \big] ^{L}  \rangle 
%\\
%& \hspace{10mm} + 4 \langle \delta \Psi ,   \big[  [ \eta \bar{\eta } \Psi ,  Q\eta \Psi  ]^{L} , \eta \Psi  \big] ^{L} \rangle 
%\end{align}

\section{Gauge-invariant insertions of picture-changing operators} %(9-14)}

In this section, we briefly review the coalgebraic description of string vertices \cite{Stasheff:1993ny, Kimura:1993ea, Getzler:2007} and how to construct NS superstring vertices \cite{Erler:2014eba}. 
%We show that using these products including insertions of picture-changing operators, one can construct a pure-gauge solution $\mathcal{G}_{L}$ of heterotic string field theory in the `small' Hilbert space and associated field. 
Since the NS string products satisfies $L_{\infty }$-relations, 
%provided that a field $\mathcal{G}$ satisfies the equation of motion of NS string field theory in the small Hilbert space, the $\mathcal{G}$-
the shifted NS string products satisfies $L_{\infty }$-relations up to the equation of motion. 

\subsection{Coalgebraic description of vertices}

Let $T(\mathcal{H} )$ be a tensor algebra of the graded vector space $\mathcal{H}$. 
As the quotient algebra of $T(\mathcal{H})$ by the ideal generated by all differences of products $A _{1}\otimes A _{2} - (-1)^{\mathrm{deg}(A _{1}) \mathrm{deg}(A _{2}) } A _{2} \otimes A _{1}$ for $A _{1}, A _{2} \in \mathcal{H}$, we can construct the symmetric algebra $S(\mathcal{H})$. 
{\it The product of states} in $S(\mathcal{H})$ is graded commutative and associative as follows
\begin{align}
A_{1}  A_{2} = (-1)^{\mathrm{deg}(A_{1}) \mathrm{deg}(A_{2}) } A_{2}  A_{1} , \hspace{5mm} A_{1}  ( A_{2}  A_{3} ) = (A_{1}  A_{2} )  A_{3} ,
\end{align}
where $A_{1},A_{2},A_{3} \in S ( \mathcal{H} )$. 
%%%%%%%%%%%%%%%%%%%%%%%%%
For us, $\mathcal{H}$ is the closed superstring state space, $S(\mathcal{H})$ is the Fock space of superstrings, and the $\mathbb{Z}_{2}$ grading, called degree, is just Grassmann parity.
{\it The product of two multilinear maps} $L:\mathcal{H}^{n} \rightarrow \mathcal{H}^{l}$, $M : \mathcal{H}^{m} \rightarrow \mathcal{H}^{k}$ also becomes a map $L \cdot M : \mathcal{H}^{n+m} \rightarrow \mathcal{H}^{k+l}$ which acts as 
\begin{align}
L \cdot M ( A _{1} A _{2} \cdots A _{n+m} ) = \sum_{\sigma } (-)^{\sigma (n,m ) }  L ( A _{\sigma (1)} \cdots A _{\sigma (n) } ) \cdot M ( A _{\sigma (n+1)} \cdots A _{\sigma (n+m) } ) .
\end{align}
Note that the $n$-product of the identity map $\mathbb{I}:\mathcal{H} \rightarrow \mathcal{H}$ on symmetric algebras is different from the $n$-tensor product or the identity $\mathbb{I}_{n}$ on $\mathcal{H}^{n}$: 
\begin{align}
\mathbb{I}_{n} \equiv  \frac{1}{n!} \overbrace{\mathbb{I}  \cdots \mathbb{I} }^{n} =  \overbrace{\mathbb{I} \otimes \cdots \otimes \mathbb{I} }^{n} .
\end{align}
In other words, $\mathbb{I}^{n}$ is the sum of all permutations, $\mathbb{I}_{n}$ gives the sum of equivalent permutations, and $\mathbb{I}_{k}\cdot \mathbb{I}_{l}$ is equivalent to the sum of different $(k,l)$-partitions of $k+l$. 

\vspace{2mm}

The $n$ string product $ L_{n} ( A_{1} A_{2} \dots A_{n} ) \equiv [A_{1} , \dots , A_{n}]$ defines a $n$-fold linear map $L_{n} : \mathcal{H}^{n} \rightarrow \mathcal{H}$. 
We can naturally define a coderivation ${\bf L}_{n} : S(\mathcal{H} ) \rightarrow  S(\mathcal{H} )$ from the map $L_{n} : \mathcal{H}^{n} \rightarrow \mathcal{H} $. 
Specifically, the coderivation ${\bf L}_{n}$ acts on the $\mathcal{H} ^{N}  \subset S(\mathcal{H} )$ as 
\begin{align}
\label{definition of bf L} 
{\bf L}_{n} A = ( L_{n} \cdot \mathbb{I} _{N-n} ) A ,  \hspace{5mm} 
{\bf L}_{n} B =0,
\end{align}
where $A \in \mathcal{H}^{N \geq n}$ and $B \in \mathcal{H}^{N < n}$. 
%%%%%%%%%%%%%%%%%%%%
Note that the commutator $\ld  {\bf L}_{m} , {\bf L}^{\prime }_{n}  \rd $ of two coderivations ${\bf L}_{m}$ and ${\bf L}_{n}^{\prime }$ also becomes a coderivation of the $(m+n-1)$-product
\begin{align}
\ld  {\bf L}_{m } , {\bf L}_{n}^{\prime }  \rd := L_{m} ( L^{\prime }_{n} \cdot \mathbb{I}_{ m-1 } ) - (-1)^{\mathrm{deg}(L_{m}) \mathrm{deg} (L^{\prime }_{n} ) } L^{\prime }_{n} ( L_{m} \cdot \mathbb{I}_{ n-1} )  .
\end{align}
Hence, in the coalgebraic description, we can write $L_{\infty }$-relations of closed string products as 
\begin{align}
\ld  {\bf L}_{1} , {\bf  L}_{n}  \rd + \ld {\bf L}_{2} , {\bf L}_{n-1} \rd + \dots + \ld  {\bf L}_{n} , {\bf L}_{1} \rd =0 ,
\end{align}
or, more simply, 
\begin{align}
\ld {\bf L} ( s ) , {\bf L} ( s ) \rd = 0 ,
\end{align}
where $s$ is a real parameter and ${\bf L}(s)$ is the generating function for the bosonic string products 
\begin{align} 
\label{generating function of bosonic string products}
{\bf L} ( s ) = \sum_{n=0}^{\infty } s^{n} {\bf L}_{n+1} .
\end{align}

\subsection{Gauge-invariant insertions}

Let $L_{N+1}^{(n)}$ be a $(N+1)$-product including $n$-insertions of picture-changing operators. 
We consider a series of these inserted products
\begin{align}
{\bf L}^{[m]}(t) := \sum_{n=0}^{\infty } t^{n} {\bf L}^{(n)}_{m+n+1} ,
\end{align}
where $t$ is a real parameter and ${\bf L} ^{(n)}_{m+n+1} $ acts on $S(\mathcal{H}) $ as (\ref{definition of bf L}). 
Note that the upper index $[m]$ on ${\bf L}^{[m]} (t)$ indicates the deficit in picture number of the products relative to what is needed for superstrings: ${\bf L}^{[0]}(t)$ is the sum of all superstring products and ${\bf L}^{[m]}(0)$ is the $(m+1)$-product of bosonic strings. 
To associate the generating functions ${\bf L}^{[0]} (t) $ of the NS superstring products with $(\ref{generating function of bosonic string products} )$, we define the following generating function
\begin{align}
{\bf L} ( s, t ) := \sum_{s=0}^{\infty } s^{m} {\bf L}^{[m]}(t)  =  \sum_{m,n=0}^{\infty } s^{m} t^{n} {\bf L}^{(n)}_{m+n+1} , 
\end{align}
where $s$ is a real parameter. 
Note that powers of $t$ count the picture number, and powers of $s$ count the deficit in picture number.
These two parameters $t$ and $s$ connect the generating function ${\bf L}(0,t) = {\bf L}^{[0]} (t)$ for the NS superstring products and the generating function ${\bf L} ( s,0)$ for the bosonic string products. 

\vspace{2mm}

The $L_{\infty }$-relations of bosonic products and derivation properties of $\eta$ can be represented by 
\begin{align}
\label{bosonic L}
\ld  {\bf L}(s,0) , {\bf L}(s,0) \rd &= 0 ,
\\
\label{bosonic small}
\ld { \pmb \eta } , { \bf L }(s,0) \rd &= 0 .
\end{align}
Starting with these relations, we can construct the NS superstring products ${\bf L} ( 0 , t )$ satisfying the $L_{\infty }$-relations and derivation properties of $\eta$, which we explain in this subsection.

\vspace{3mm}

\underline{Gauge-invariant insertions}

\vspace{2mm}

To construct ${\bf L} ( 0 , t ) $ satisfying the $L_{\infty }$-relations and $\eta$-derivation conditions
\begin{align}
\label{super L}
\ld  {\bf L}( 0, t) , {\bf L} (0,t) \rd &= 0 ,  
\\
\label{super small}
\ld   {\pmb \eta } , {\bf L} ( 0 , t )  \rd &= 0 ,
\end{align}
we have to solve the following pair of differential equations 
\begin{align}
\label{L equation}
\frac{\partial }{\partial t} {\bf L} (s,t) &= \ld {\bf L} (s,t) , {\bf \Xi } (s,t) \rd ,
\\ 
\label{small equation}
\frac{\partial }{\partial s} {\bf L} (s,t) &= \ld  {\pmb \eta } , {\bf \Xi }(s,t) \rd  , 
\end{align}
with the initial conditions $(\ref{bosonic L})$ and $(\ref{bosonic small})$ at $t=0$. 
As well as ${\bf L} ( s , t )$, we define a generating function ${\bf \Xi } ( s, t )$ for gauge parameters 
\begin{align}
{\bf \Xi } (s,t) := \sum_{m=0}^{\infty } s^{m} {\bf \Xi }^{[m]} (t) = \sum_{m,n=0}^{\infty } s^{m} t^{n} {\bf \Xi }^{(n+1)}_{m+n+2} .
\end{align}
The solution of this pair of differential equations generates all products including appropriate insertions of picture-changing operators. 

\vspace{3mm}

We can obtain the superstring $L_{\infty}$-relations $( \ref{super L} )$ as a solution of the differential equation 
\begin{align}
\label{LL}
\frac{\partial }{\partial t } \ld {\bf L}( s, t ) , {\bf L} ( s , t ) \rd &= \Ld \ld  {\bf L}( s,t ) , {\bf L} ( s,t ) \rd   , {\bf \Xi }  ( s , t )  \Rd  ,  
\end{align}
where ${\bf \Xi } ( s ,t )$ is a generating function for gauge parameters, if we impose the initial condition $( \ref{bosonic L} )$ at $t=0$. 
Provided that ${\bf L} ( s,t ) $ satisfies $( \ref{L equation} )$, the equation $(\ref{LL})$ automatically holds because of Jacobi identities $\Ld \ld  {\bf L}( s,t ) , {\bf L} ( s,t ) \rd   , {\bf \Xi }  ( s , t )  \Rd = 2 \Ld \ld {\bf L} (s,t) , {\bf \Xi }(s,t) \rd , {\bf L}(s,t) \Rd  $.

Further, the equation $(\ref{L equation})$ leads to the equation 
\begin{align}
\frac{\partial }{\partial t} \ld {\pmb \eta } , {\bf L} (s,t) \rd  = \Ld \ld {\pmb \eta } , {\bf L} ( s, t ) \rd , {\bf \Xi } ( s,t ) \Rd + \Ld \ld {\pmb \eta } , { \bf  \Xi } (s ,t ) \rd , {\bf L}( s , t )   \Rd .
\end{align}
%when ${\bf L} (s ,t )$ satisfies $(\ref{L equation})$.  
Therefore, when ${\bf L} ( s, t ) $ satisfies the equation $(\ref{small equation})$, we obtain 
\begin{align}
\frac{\partial }{\partial t} \ld {\pmb \eta } , {\bf L}(s,t)  \rd = \Ld  \ld {\pmb \eta } , {\bf L} (s,t ) \rd , {\bf \Xi } (s,t)  \Rd  -\frac{1}{2}\frac{\partial }{\partial s} \ld  {\bf L}(s,t) , {\bf L} (s,t) \rd   
\end{align}
and the differential equation $\frac{\partial }{\partial t} \ld {\pmb \eta } , {\bf L} (s,t) \rd =  \Ld  \ld {\pmb \eta } , {\bf L} (s,t ) \rd , {\bf \Xi } (s,t)  \Rd$ (up to $\ld  {\bf L}(s,t) , {\bf L} (s,t) \rd$) with the initial condition $(\ref{bosonic small})$ indicates the $\eta$-dirivative conditions $(\ref{super small})$ of the superstring products. 
As a result, the pair of equations $(\ref{L equation})$ and $(\ref{small equation})$ generates inserted products ${\bf L}$ satisfying the $L_{\infty }$-relations $\ld  {\bf L} (s,t) , {\bf L} (s,t) \rd =0 $ and $\eta $-derivative conditions $\ld  {\pmb \eta } , {\bf L} (s,t) \rd =0 $. 

\vspace{3mm}

Expanding $(\ref{L equation})$ and $(\ref{small equation})$ in powers of $(s,t)$, we obtain the following formulae 
\begin{align}
\label{expanding L}
{\bf L} ^{(n+1)}_{m+n+2}  = & \frac{1}{n+1} \sum_{k=0}^{n} \sum_{l=0}^{m} \Ld  {\bf L }^{(k)}_{k+l+a} , {\bf \Xi }^{(n-k+1)}_{m+n+2-k-l}  \Rd  , 
\\ 
\label{primitive}
\ld {\pmb \eta } , {\bf \Xi } ^{(n+1)}_{m+n+2}  \rd = & (m+1) {\bf L}^{(n)}_{m+n+2} , 
\end{align}
at each order of $s^{m}t^{n}$ and these formulae determine $L_{m+n+2}^{(n)}$ and $\Xi _{m+n+2}^{(n+1)}$ recursively. 

Note that ${\bf \Xi } ( s , t)$ is not unique, however, there exists the one including symmetric insertions 
\begin{align}
\label{left chiral construction}
{\bf \Xi }^{(n+1)}_{m+n+2} = \frac{m+1}{m+n+3} \Big(  \xi {\bf L}^{(n)}_{m+n+2} - {\bf L}^{(n)}_{m+n+2} \big( \xi \cdot \mathbb{I}^{m+n+1} \big) \Big) .
%\sum_{k=0}^{m+n+1} \mathbb{I}^{\otimes k} \otimes \xi \otimes \mathbb{I}^{\otimes (m+n+1-k)}  \Big)
\end{align}
Therefore, we can always derive explicit forms of these inserted products as follows: 
\begin{align}
\nonumber 
{\bf L}^{(0)}_{n+1} &= \mathrm{given } ,
\\ \nonumber 
\Xi ^{(1)}_{n+1} &= \frac{n}{n+2} \Big(  \xi {\bf L}^{(0)}_{n+1} - {\bf L}^{(0)}_{n+1} \big( \xi \cdot \mathbb{I}^{n} \big) \Big) ,
\\ \nonumber 
{\bf L}^{(1)}_{n+1} &= {\bf \ld Q} , {\bf \Xi }^{(1)}_{n+1} \rd + \ld {\bf L}^{(0)}_{2} , {\bf \Xi }^{(1)}_{n} \rd + \dots + \ld {\bf L }^{(0)}_{n} , {\bf \Xi }^{(1)}_{2} \rd  ,
\\ \nonumber 
\Xi ^{(2)}_{n+1} &= \frac{n-1}{n+2} \Big(  \xi {\bf L}^{(1)}_{n+1} - {\bf L}^{(1)}_{n+1} \big( \xi \cdot \mathbb{I}^{n} \big) \Big) ,
\\ \nonumber 
{\bf L}^{(2)}_{n+1}&= \frac{1}{2} \Big( {\bf \ld Q} , {\bf \Xi }^{(2)}_{n+1} \rd + \ld {\bf L}^{(0)}_{2} , {\bf \Xi }^{(2)}_{n} \rd + \ld  {\bf L}^{(1)}_{2} , {\bf \Xi }^{(1)}_{n} \rd + \dots ,
\\ \nonumber 
& \hspace{15mm} + \ld  {\bf L}^{(0)}_{n-1} , {\bf \Xi }^{(2)}_{3}  \rd + \ld {\bf L}^{(1)}_{n-1} , {\bf \Xi }^{(1)}_{3} \rd + \ld {\bf L }^{(1)}_{n} , {\bf \Xi }^{(1)}_{2} \rd   \Big) ,
\\ \nonumber 
&\vdots
\\ \nonumber 
\Xi ^{(n)}_{n+1}&=\frac{1}{n+2} \Big(  \xi {\bf L}^{(n)}_{n+1} - {\bf L}^{(n)}_{n+1} \big( \xi \cdot \mathbb{I}^{n} \big) \Big) ,
\\
{\bf L}^{(n)}_{n+1} &= \frac{1}{n} \Big( {\bf \ld Q} , {\bf \Xi }^{(n)}_{n+1} \rd + \ld {\bf L}^{(1)}_{2} , {\bf \Xi }^{(n-1)}_{n} \rd + \dots + \ld {\bf L }^{(n-1)}_{n} , {\bf \Xi }^{(1)}_{2} \rd   \Big) .
\end{align}
For example, we find that the lowest inserted product ${\bf L}^{(1)}_{2}$ is given by
\begin{align}
\nonumber 
{\bf L}_{2}^{(1)} ( A , B ) & = \ld  {\bf Q} , {\bf \Xi }^{(1)}_{2} \rd  ( A , B ) 
\\ 
& = \frac{1}{3}\Big(  X [ A  , B ] + [ X A , B ] + [ A , X B ] \Big)  ,
\end{align}
where $L_{2}^{(0)}( A , B ) = [ A , B ]$, and the second lowest product ${\bf L}^{(2)}_{3}$ is given by 
\begin{align}
{\bf L}^{(2)}_{3} ( A , B , C ) = \frac{1}{2} \Big(  \ld {\bf Q} , {\bf \Xi }^{(2)}_{3} \rd + \ld {\bf L}^{(1)}_{3} , {\bf \Xi }^{(1)}_{3} \rd  \Big)  (A,B,C) ,
\end{align}
where $L_{3}^{(0)} (A,B,C)=[A,B,C]$ and 

\begin{align}
\nonumber 
\frac{1}{2} \ld {\bf Q} , {\bf \Xi }^{(2)}_{3} \rd ( A, B, C ) %= \, & \frac{1}{8} \Big(  X^{2} [A,B,C] + [X^{2} A ,B ,C ,] + [A, X^{2} B , C, ] + [A,B,X^{2}C] \Big) 
%\\ \nonumber 
%&- \frac{1}{8} \Big(  \xi [ [ X A, B ] + [ A, X B ] , C ] + \xi [ [ A, B ] , X C ]  \Big) 
%\\ \nonumber 
%&- \frac{1}{8} \Big(  X [ [ \xi A, B ] + [ A, \xi B ] , C ] + X [ [ A, B ] , \xi C ]   \Big) 
%\\ \nonumber 
%&+\frac{1}{12} \Big(   X [ \xi [ A,B ] , C] + [ \xi [ X A , B ] + \xi [ A ,X B ] , C ] + [ \xi [ A ,B ] , X C ] \Big) 
%\\ \nonumber
%&+\frac{1}{12} \Big(   \xi [ X [ A,B ] , C] + [ X [ \xi A , B ] + X [ A , \xi B ] , C ] + [ X [ A ,B ] , \xi C ] \Big) 
%
%\\
%& \hspace{5mm} + \big{\{ } ( B, C,A )  { \mathchar`- }  \mathrm{terms} \big{\} }+ \big{\{ } (C,A,B ) { \mathchar`-}  \mathrm{terms} \big{\} }  , 
%%%%%%%%%%%%%
= \, & \frac{1}{8\cdot 2} \Big{\{ } X^{2} \big[ A,B,C \big] + [X^{2} A ,B ,C ,] + [A, X^{2} B , C, ] + [A,B,X^{2}C]  \Big{\} } 
\\ \nonumber 
& \hspace{3mm}  + \frac{1}{8 \cdot 2} \Big{\{ } \xi X [ [  A, B ] , C]  + [ [ \xi X A,  B ] , C ] 
\\ \nonumber 
& \hspace{20mm} + (-)^{A} [ [ A, \xi X B ] ,  C ] + (-)^{A+B} [ [ A ,B ] , \xi X C ] \Big{\} } 
\\ \nonumber 
& \hspace{3mm} + \frac{1}{8} \Big{\{ }  X[ X A,B,C] + X[A, XB,C] + X[ A,B,XC] 
\\ \nonumber 
& \hspace{20mm} + [XA,XB,C] +[XA,B,XC] + [A,XB,XC]    \Big{\} }  
\\ \nonumber 
& \hspace{3mm} - \frac{1}{8\cdot 2} \Big{\{ }  \xi [ [ X A, B ] + [ A, X B ] , C ] + \xi [ [ A, B ] , X C ] 
\\ \nonumber 
& \hspace{13mm} +  X [ [ \xi A, B ] , C ]  -  [ [ \xi A, X B ] ,C]  -  [ [ \xi A, B ] , X C ]  
\\ \nonumber 
& \hspace{13mm}  + (-)^{A} \Big( X[ [ A, \xi B ] , C ] -  [ [ X A , \xi B ] , C ] - [ [ A , \xi B ] , X C ] \Big) 
\\ \nonumber 
& \hspace{13mm} + (-)^{A+B}  \Big( X [ [ A, B ] , \xi C ]  -  [ [ X A , B ] , \xi C ] - [ [ A, X B ] , \xi C ] \Big)  \Big{\} } 
\\ \nonumber 
& \hspace{3mm} + \frac{1}{4\cdot 3} \Big{\{ }  X [ \xi [ A,B ] , C] + [ \xi [ X A , B ] + \xi [ A ,X B ] , C ] + [ \xi [ A ,B ] , X C ]  
\\ \nonumber 
& \hspace{20mm} + \xi \big[ X [ A,B ] , C \big] + \big[ X [ \xi A , B ] , C \big]  
\\ \nonumber 
& \hspace{25mm} + (-)^{A}  \big[  X [ A , \xi B ]  , C \big]  + (-)^{A+B} \big[ X [ A ,B ] , \xi C \big] \Big{\} } 
\\ 
& \hspace{15mm} + \big{\{ }  ( B, C ,A ) {\mathchar`-} \mathrm{terms}  \big{\} } + \big{\{ }  (  C ,A , B ) \mathchar`- \mathrm{terms} \big{\} } .
%%%%%%%%%%%%%
\end{align}
\begin{align}
\nonumber 
\frac{1}{2} \ld {\bf L}^{(1)}_{2} , {\bf \Xi }^{(1)}_{2}  \rd ( A, B, C ) = 
%\, & \frac{1}{9 \cdot 2} \Big(   X [ \xi [ A,B ] , C] + [ \xi [ X A , B ] + \xi [ A ,X B ] , C ] + [ \xi [ A ,B ] , X C ] 
%\\ \nonumber 
%& - \xi [ X [ A,B ] , C] - [ X [ \xi A , B ] + X [ A , \xi B ] , C ] - [ X [ A ,B ] , \xi C ]  
%\\  \nonumber 
%&+ 2 [ \xi X [ A ,B ] , C ]  
%\\ \nonumber 
%& - X [ [ \xi A , B ] + [ A , \xi B ] , C ] - [ [ \xi A, B ] + [ A , \xi B ] , X C ]  
%\\ \nonumber 
%& - \xi [ [ X A , B ] + [ A , X B ] , C ] - [ [ X A , B ] + [ A , X B ] , \xi C ]  \Big) 
%
%\\
%& \hspace{5mm} + \big{\{ } ( B, C,A )  { \mathchar`- }  \mathrm{terms} \big{\} }+ \big{\{ } (C,A,B ) { \mathchar`-}  \mathrm{terms} \big{\} } .
%%%%%%%%%%
 \, & \frac{1}{9 \cdot 2} \Big{\{ } 2 \big[ \xi X [ A ,B ] , C \big]  - \xi \big[ X [ A,B ] + [ X A , B ] + [ A , X B ] , C \big]  
\\ \nonumber 
& \hspace{7mm} + X \big[ \xi [ A,B ] , C \big] + \big[ \xi \big( [ X A , B ] + [ A ,X B ] \big)  , C \big] + \big[ \xi [ A ,B ] , X C \big] 
\\ \nonumber  
& \hspace{9mm} - X \big[ [ \xi A , B ] , C \big]  - \big[ X  [ \xi A , B ] , C \big]  - \big[ [ \xi A, B ] , X C \big]  
\\ \nonumber 
& \hspace{7mm} - (-)^{A} \Big( \big[  X [ A , \xi B ] , C \big] + X \big[ [ A , \xi B ] , C \big] + \big[ [ A , \xi B ] , X C \big] \Big) 
\\ \nonumber 
& \hspace{9mm} - (-)^{A+B} \Big( \big[ X [ A ,B ]  + [ X A , B ] + [ A , X B ] , \xi C \big] \Big) \Big{\} } 
\\ 
& \hspace{5mm} + \big{\{ } ( B, C,A )  { \mathchar`- }  \mathrm{terms} \big{\} }+ \big{\{ } (C,A,B ) { \mathchar`-}  \mathrm{terms} \big{\} } 
\end{align}

%\clearpage

\subsection{NS string products}  %{Useful properties}

The generating function ${\bf L}( 0 , t )$ of the superstring products, as well as that of bosonic ones ${\bf L} (s,0) $, has nice properties, which we explain in this subsection. 
Note that the above ${\bf L}_{n+1}^{(n)}$ obtained from $(\ref{expanding L})$ and $(\ref{left chiral construction})$ carries ghost-and-picture number $(1-2n | n,0 )$. 
In this sense, we write ${\bf L}_{n+1}^{(n,0)}$ for this ${\bf L}_{n+1}^{(n)}$, an NS superstring product with insertions of {\it left}-moving picture-changing operators. 
${\bf L}_{n+1}^{(n , 0)}$ gives the $(n+1)$-product of NS (heterotic) superstring field theory in the {\it small} Hilbert space of {\it left} movers \cite{Erler:2014eba}. 

By construction, we can also obtain an NS superstring product ${\bf L}_{n+1}^{(0,n)}$ with insertions of {\it right}-moving picture-changing operators $\widetilde{X}$ by replacing $(\eta , \xi , X)$ with $(\tilde{\eta } , \tilde{\xi } , \widetilde{X})$ in $(\ref{expanding L})$, $(\ref{primitive})$, and $(\ref{left chiral construction})$. 
In the rest, we consider these {\it right}-mover NS products $\{ {\bf L}_{n+1}^{(0,n)} \} _{n=0}^{\infty }$ and write 
\begin{align}
[A_{0}, A_{1} , \dots , A_{n}]^{L} := {\bf L}^{(0,n)}_{n+1} (A_{0} , A_{1} ,\dots , A_{n} ) . 
\end{align}
The right-mover NS products also satisfies $L_{\infty }$-relations 
\begin{align}
\sum_{\sigma} \sum_{k} (-1)^{|\sigma ( A )|} \big[  [  A_{\sigma (1)} , \dots , A_{\sigma (k)} ]^{L} , A_{\sigma (k+1) } , \dots , A_{\sigma (n)} \big] ^{L} = 0 .
\end{align}
Note also that the $n$-product $[ A_{1} , \dots , A_{n} ]^{L}$ has ghost-and-picture number $( 3-2n | 0 , n-1 )$.

\vspace{3mm}

\underline{$L_{\infty }$-properties of right-mover NS string products}

\vspace{2mm}

Let $\mathcal{G}$ be a ghost-and-picture number $(2|0,-1)$ state and $A, A_{1}, \dots , A_{n}$ be arbitrary states. 
We can define a shifted BRST operator $Q_{\mathcal{G}}$ and shifted right-mover NS string products 
\begin{align}
Q_{\mathcal{G}} A \equiv [  A \, ]_{\mathcal{G}}^{L} :=& \,\, Q A + \sum_{n=1}^{\infty }  \frac{ \kappa ^{n} }{ n! }  [ \, \mathcal{G} ^{n} , \, A \, ] ^{L} ,
\\
[ A_{1} , \dots , A_{n} ] _{ \mathcal{G}} ^{L} :=& \sum_{k=1}^{\infty } \frac{ \kappa ^{k} }{ k! } [  \mathcal{G} ^{k}, A_{1} , \dots , A_{n}  ]^{L} ,
\end{align}
in the same manner as shifted bosonic string products. 
Provided that the state $\mathcal{G}$ shifting these products satisfies the equation of motion $\mathcal{F}(G) =0 $ of NS (heterotic) string field theory in the small Hilbert space of right movers 
\begin{align}
\mathcal{F} (\mathcal{G} ) \equiv Q \mathcal{G} + \sum_{n=1}^{\infty } \frac{\kappa ^{n} }{(n+1)!} \big[  \mathcal{G}^{n} , \mathcal{G} \big] ^{L} = 0 , 
\end{align}
these shifted products satisfy (strong) $L_{\infty }$-relations\footnote{
For general $\mathcal{G}$, weak $L_{\infty }$-relations hold, which are equivalent to (strong) $L_{\infty }$-relations up to $\mathcal{F} ( \mathcal{G} )$.}: 
\begin{align}
\sum_{\sigma }\sum_{k} (-1)^{ |\sigma | }\big{[}  [ A_{\sigma (1)} , \dots , A_{\sigma (k) }  ]^{L}_{\mathcal{G}}  , A_{\sigma (k+1 )} , \dots , A_{\sigma (n )}   \big{]} ^{L}_{\mathcal{G}} =0 .
\end{align}
Then, $Q_{\mathcal{G}}$ becomes a nilpotent operator.

\section{WZW-like expression}% (15-19)}

In this section, first, we gives the defining equations of a formal pure-gauge $\mathcal{G}_{L}$ and associated fields $\Psi _{t}, \Psi _{\eta }, \Psi _{\delta }$. 
These are functions of NS-NS string fields $\Psi $ and become key ingredients of our construction. 
Then, we present a closed form expression of WZW-like action for NS-NS string field theory, the equation of motion, and the gauge invariance of the action.

\subsection{Pure-gauge $\mathcal{G}_{L}$ and `large' associated field $\Psi _{\mathbb{X}}$}  %{Useful properties}

The NS-NS string field $\Psi $ is a Grassmann-even and ghost-and-picture number $(0|0,0)$ state living in the large Hilbert space of left-and-right movers: $\eta \Psi \not= 0$ and $\tilde{\eta }\Psi \not= 0$. 

\vspace{3mm}

\underline{A pure-gauge $\mathcal{G}_{L}$ of right-mover NS theory}

\vspace{2mm}

We can build a formal pure-gauge solution $\mathcal{G}_{L}$ of NS heterotic string field theory in the small Hilbert space of right-movers with a finite gauge parameter $\tilde{\eta } \Psi $ living in the left-mover large and right-mover small Hilbert space by successive infinitesimal gauge transformations. 
The pure-gauge field $\mathcal{G}_{L} = \mathcal{G}_{L} [ \tilde{\eta } \Psi  ]$ is a function of $\tilde{\eta } \Psi $, defined by the $\tau =1$ solution of the differential equation
\begin{align}
\nonumber 
\label{definition of g}
\frac{\partial }{\partial \tau } \mathcal{G}_{L} [ \tau \tilde{\eta } \Psi ]  &= Q \, \tilde{\eta } \Psi  + \sum_{n=1}^{\infty } \frac{\kappa ^{n}}{n!} \big[ \big( \mathcal{G}_{L}[\tau \tilde{\eta } \Psi ] \big) ^{n} , \tilde{\eta } \Psi \big] ^{L} 
\\ 
& \equiv  \, Q_{\mathcal{G}_{L} [ \tau \tilde{\eta } \Psi ] }  \tilde{\eta } \Psi    
\end{align}
with the initial condition $\mathcal{G}_{L} [ 0 ] =0$, where $\tau \in [0,1]$ is a real parameter connecting $0$ and $\mathcal{G}_{L} [ \tilde{\eta } \Psi ]$. 
(See also \cite{Schubert:1991en, Berkovits:2004xh}, Appendix A, and Appendix B.) 
Solving the defining equation $(\ref{definition of g})$ and setting $\tau =1$,  we obtain the explicit form of the pure-gauge $\mathcal{G}_{L} \equiv \mathcal{G}_{L} [\tau = 1]$ as follows 
\begin{align}
\mathcal{G}_{L} =  Q \tilde{\eta } \Psi + \frac{\kappa }{2}  [ Q \tilde{\eta } \Psi , \tilde{\eta } \Psi ]^{L} + \frac{\kappa ^{2}}{3!} \Big(  \big[ Q \tilde{\eta } \Psi , Q \tilde{\eta } \Psi , \tilde{\eta } \Psi \big] ^{L} + \big[ [ Q \tilde{\eta } \Psi , \tilde{\eta } \Psi ]^{L}  \tilde{\eta } \Psi \big] ^{L} \Big) + \dots   .
\end{align}

\vspace{2mm}

Note that $\mathcal{G}_{L} $ is  a Grassmann even and ghost-and-picture number $(2|0,-1)$ state satisfying $ \eta \mathcal{G}_{L} \not= 0$ and $\tilde{\eta } \mathcal{G}_{L} = 0$, the field $\tilde{\eta } \Psi$ is a Grassmann odd and ghost-and-picture number $(1|0, -1)$ state satisfying $\eta ( \tilde{\eta } \Psi ) \not= 0$ and $\tilde{\eta } (\tilde{\eta } \Psi ) = 0$, and the $n$-product $[A_{1}, \dots , A_{n}]^{L}$ is a Grassmann odd and ghost-and-picture number $( 3-2n | 0, n-1 )$ product satisfying $\Delta _{\eta / \tilde{\eta }} [A_{1} , \dots , A_{n} ]^{L} = 0$.

\vspace{3mm}

\underline{An associated field $\psi _{\mathbb{X}}$ with derivation $\mathbb{X}$}

\vspace{2mm}

In the rest, we simply write $\mathcal{G}_{L} [\tau ]$ for the intermediate pure-gauge field rather than $\mathcal{G}_{L} [ \tau \tilde{\eta } \Psi ]$. 
There exists a special string field $\psi _{\mathbb{X}}$, so-called {\it an associated field}, satisfying  
\begin{align}
(-1)^{\mathbb{X}} \mathbb{X} \, \mathcal{G}_{L} = Q_{\mathcal{G}_{L} } \psi _{\mathbb{X}}  ,
\end{align}
where $\mathbb{X}$ is a derivation of our right-mover $(-,\mathrm{NS})$ string products $[A_{1} , \dots , A_{n} ]^{L}$: 
\begin{align}
(-1)^{\mathbb{X}} \mathbb{X} [ A_{1} , \dots , A_{n} ]^{L} + \sum_{i=1}^{n} (-1)^{ \mathbb{X} (A_{1} + \dots +A_{i-1}) } [ A_{1} , \dots , \mathbb{X} A_{i} , \dots A_{n} ]^{L}=0.
\end{align} 
Utilizing the $\mathcal{G}_{L}[\tau ]$-shifted two-product $[ A_{1} , A_{2} ]^{L}_{\mathcal{G}_{L}[\tau ]}$, the defining equation of $\psi _{\mathbb{X}} $ is given by 
\begin{align}
\label{associated field}
\frac{\partial }{\partial \tau } \psi _{\mathbb{X}} [\tau ] = \mathbb{X} \, \tilde{\eta } \Psi + \kappa \big[  \tilde{\eta } \Psi ,  \psi _{\mathbb{X}} [\tau ] \big] ^{L}_{\mathcal{G}_{L} [\tau ]} 
\end{align}
with the initial condition $\psi _{\mathbb{X}} [0] =0$. 
Note that $\psi _{\mathbb{X}} [ \tau ]$, as well as $\mathcal{G}_{L} [ \tau ]$, is a function of $\tau \tilde{\eta } \Psi$ and $\tau $ is a real parameter connecting $0$ and $\psi _{\mathbb{X}} := \psi _{\mathbb{X} } [ 1 ]$. 
The associated filed $\psi _{\mathbb{X}}$ carries ghost-and-picture number $( {\bf G}_{\mathbb{X} } + 1|  {\bf p}_{\mathbb{X}} , \tilde{{\bf p}}_{\mathbb{X}} -1 )$, where $({\bf G}_{\mathbb{X}}  | {\bf p}_{\mathbb{X}}  ,  \tilde{{\bf p}}_{\mathbb{X}})$ is that of $\mathbb{X}$.

\vspace{3mm}

\underline{A `large' associated field $\Psi _{\mathbb{X}}$}

\vspace{2mm}

These pure-gauge field $\mathcal{G}_{L}$ and associated field $\psi _{\mathbb{X}}$ belong to the left-mover large and right-mover small Hilbert space: $\eta \, \mathcal{G}_{L} \not= 0$, $\eta \, \psi _{\mathbb{X} } \not= 0$, and $\tilde{\eta } \, \mathcal{G}_{L} =\tilde{\eta } \, \psi _{ \mathbb{X} } =0$. 
Since $\tilde{\eta }$-cohomology is trivial in the large Hilbert space of left-and-right movers, there exist large fields $G_{L}$ and $\Psi _{ \mathbb{X} }$ such that 
\begin{align}
\mathcal{G}_{L} = \tilde{\eta } \, G_{L},  \hspace{5mm}  \psi _{\mathbb{X}}  =  \tilde{\eta } \, \Psi _{\mathbb{X}} .
\end{align}
Note that the relation $ \tilde{\eta } \, \mathbb{X} G_{L}= - \tilde{\eta } \, Q_{\mathcal{G}_{L}} \Psi _{\mathbb{X}} $ holds because of $(-)^{\mathbb{X}} \mathbb{X} (\eta G_{L} ) = Q_{\mathcal{G}_{L}} ( \tilde{\eta } \Psi _{\mathbb{X}} ) $ and $\tilde{\eta } \mathcal{G}_{L} =0 $. %the following relation 
Hence, up to $Q_{\mathcal{G}_{L}}$- and $\tilde{\eta }$-exact terms, these large-space fields $G_{L}$ and $\Psi _{\mathbb{X}}$ satisfy 
\begin{align}
\mathbb{X} \, G_{L} = - Q_{\mathcal{G}_{L} } \Psi _{ \mathbb{X} } , 
\end{align}
and the defining equations (up to $Q_{\mathcal{G}_{L}}$- and $\tilde{\eta }$-exact terms) of $G_{L}$ and $\Psi _{\mathbb{X}}$ are given by
\begin{align}
\label{large G}
&\frac{\partial }{\partial \tau } G_{L} [\tau ] = -  Q_{\mathcal{G}_{L} [\tau ] } \Psi , 
\\ % \hspace{5mm} 
\label{large associated field}
&\frac{\partial }{\partial \tau } \Psi _{ \mathbb{X} } [\tau ] = (-1)^{ \mathbb{X} } \mathbb{X} \Psi + \kappa \big[ \tilde{\eta } \Psi , \Psi _{ \mathbb{X} } [\tau ] \big] ^{L}_{\mathcal{G}_{L} [ \tau ] } ,
\end{align}
with the initial conditions $G_{L}[\tau =0 ] = 0$ and $\Psi _{\mathbb{X}} [ \tau =0 ] =0$. 
As well as $\mathcal{G}_{L}$ and $\psi _{\mathbb{X}}$, large fields $G_{L}$ and $\Psi _{\mathbb{X}}$ are also functions of $(\Psi , \tilde{\eta } \Psi )$. 
Here $\tau $ is a real parameter connecting $0$ and $\tilde{\eta } \Psi $. 
With the initial condition $\Psi _{\mathbb{X} } [\tau = 0] = 0$, we find that the first few terms in $\Psi _{\mathbb{X}}$ are given by 
\begin{align}
(-1)^{\mathbb{X} } \Psi _{\mathbb{X} } = \mathbb{X} \Psi + \frac{\kappa }{2} [ \tilde{\eta } \Psi , \mathbb{X}  \Psi ]^{L} + \frac{\kappa ^{2}}{3!} \Big(  2 [ \tilde{\eta } \Psi , Q \tilde{\eta } \Psi , \mathbb{X} \Psi] ^{L} + [ \tilde{\eta } \Psi , [ \tilde{\eta } \Psi , \mathbb{X} \Psi ]^{L} ]^{L}  \Big) + \dots .
\end{align}
Note that the large associated field $\Psi _{\mathbb{X}}$ has the same ghost-and-picture number as $\mathbb{X}$. 

\vspace{3mm}

\underline{A $t$-parametrized large field $\Psi _{t} $}

\vspace{2mm}

Let $\Psi (t)$ be a $t$-parametrized path connecting $\Psi (0) =0$ and $\Psi (1) =\Psi $. 
The above defining equations of $\mathcal{G}_{L}$, $\psi _{\mathbb{X}}$, $G_{L}$, and $\Psi _{\mathbb{X}}$ hold not only for the field $\Psi$ but for the $t$-parametrized field $\Psi (t)$. 
Hence, we can built $t$-parametrized ones $\mathcal{G}_{L}(t)$, $\psi _{\mathbb{X}} (t)$, $G_{L} (t)$, and $\Psi _{\mathbb{X}} (t)$ by replacing $\Psi $ with $\Psi (t)$ in the defining equations of $\mathcal{G}_{L}$, $\psi _{\mathbb{X}}$, $G_{L}$, and $\Psi _{\mathbb{X}}$. 
For example, for $\mathbb{X}= \partial _{t}$, solving (\ref{large associated field}) with replacement of $\Psi $ and setting $\tau =1$, we obtain the $t$-parametrized field $\Psi _{t} \equiv \Psi _{\partial_{t}} (t)$
\begin{align}
\label{3.3-t}
\nonumber 
\Psi _{ t } =  \partial _{t} \Psi (t) + \, & \frac{\kappa }{2} [ \tilde{\eta } \Psi (t) , \partial _{t}  \Psi (t) ]^{L} 
+ \frac{\kappa ^{2}}{3!} \Big(  2 [ \tilde{\eta } \Psi (t) , Q \tilde{\eta } \Psi (t) , \partial _{t} \Psi (t) ] ^{L} 
\\
& \hspace{10mm}  + [ \tilde{\eta } \Psi (t) , [ \tilde{\eta } \Psi (t) , \partial _{t} \Psi (t) ]^{L} ]^{L}  \Big) + \dots ,
\end{align}
which appears in the action for NS-NS string fields with general $t$-parametrization. 
Note that this $\Psi _{t}$ has the same ghost-and-picture number $(0|0,0)$ as NS-NS string field $\Psi $, and the equation $\tilde{\eta } \Psi _{t} = \tilde{\eta } \Psi $ holds for the linear path $\Psi (t) = t \Psi $.

\subsection{Wess-Zumino-Witten-like action}%{All vertices as the $\mathcal{G}_{L}$-shifted products}

Let $\mathcal{G}_{L} = \sum_{n=0}^{\infty } \kappa ^{n} \mathcal{G}_{L}^{(n)}$ be the expansion of the pure-gauge $\mathcal{G}_{L}$ in powers of $\kappa $. 
Here, we propose a large-space WZW-like action utilizing the pure-gauge $\mathcal{G}_{L}(t)$ and the large associated field $\Psi _{t}$.

\vspace{3mm}

\underline{The generating function for $V_{n} ( \Psi ^{n} )$}

\vspace{2mm}

Recall that the kinetic term $S_{2} = \frac{1}{2} \langle \Psi , V_{1} ( \Psi ) \rangle $ is given by $V_{1} (\Psi ) = \eta Q  \tilde{\eta } \Psi $, which is equivalent to $\eta \mathcal{G}_{L}^{(0)} $. 
In section 2, we derived the gauge-invariant cubic vertex $V_{2}$ of $S_{3} = \frac{1}{3!} \langle \Psi , V_{2} ( \Psi ^{2} ) \rangle $
\begin{align}
V_{2} (\Psi ^{2}) = 
\eta [ Q \tilde{\eta } \Psi , \tilde{\eta } \Psi ]^{L} =& \,\, \frac{\eta }{3!} \Big( \widetilde{X} [ Q \tilde{\eta } \Psi , \tilde{\eta } \Psi ] + [ \widetilde{X} Q \tilde{\eta } \Psi , \tilde{\eta } \Psi ] + [ Q \tilde{\eta } \Psi , \widetilde{X} \tilde{\eta } \Psi ] \Big) ,
\end{align}
and the gauge-invariant quartic vertex $V_{3}$ of $S_{4} = \frac{1}{4!} \langle \Psi , V_{3} ( \Psi ^{3} ) \rangle $
\begin{align}
V_{3} (\Psi ^{3} ) = 
\eta \Big( \big[ Q \tilde{\eta } \Psi , Q \tilde{\eta } \Psi , \tilde{\eta } \Psi  \big] ^{L} + \big[  [ Q \tilde{\eta } \Psi , \tilde{\eta } \Psi ]^{L} , \tilde{\eta } \Psi \big] ^{L}  \Big) ,
\end{align}
which are equivalent to $2 \cdot \eta \, \mathcal{G}_{L}^{(1)}$ and $3! \cdot \eta \, \mathcal{G}_{L}^{ (2)}$ respectively. 
Note that quintic vertex 
\begin{align}
\nonumber 
V_{4} (\Psi ^{4} ) = & \eta \Big(  \big[  ( Q \tilde{\eta } \Psi )^{3} , \tilde{\eta } \Psi \big] ^{L}  +  \big[  [ ( Q \tilde{\eta } \Psi )^{2} , \tilde{\eta } \Psi  ]^{L} , \tilde{\eta } \Psi  \big] ^{L} 
\\
& \hspace{7mm} + 3 \big[  [ Q \tilde{\eta } \Psi  ,  \tilde{\eta } \Psi  ]^{L} , Q \tilde{\eta } \Psi , \tilde{\eta } \Psi \big] ^{L}  + \big[ [ [ Q \tilde{\eta } \Psi , \tilde{\eta } \Psi ]^{L} , \tilde{\eta } \Psi ]^{L}  \big] ^{L} \Big) ,
\end{align}
is also given by $4! \cdot \eta \, \mathcal{G}_{L }^{ (3)}$. 
Similarly, the relation $V_{n} ( \Psi ^{n} ) = n! \cdot \eta \, \mathcal{G} _{L}^{(n-1)}$ holds for the $(n+1)$-point vertex $V_{n}$. 
Therefore, the pure-gauge solution $\mathcal{G}_{L}$ 
\begin{align}
\mathcal{G}_{L} = Q \tilde{\eta } \Psi + \frac{\kappa }{2} [ Q \tilde{\eta } \Psi , \tilde{\eta } \Psi ]^{L} + \frac{\kappa ^{2}}{3!} \Big(  [ Q \tilde{\eta } \Psi ,Q \tilde{\eta } \Psi , \tilde{\eta } \Psi ]^{L} + \big[  [ Q \tilde{\eta } \Psi , \tilde{\eta } \Psi ]^{L} , \tilde{\eta } \Psi \big] ^{L} \Big) + \dots   ,
\end{align}
defined by $\partial _{\tau } \mathcal{G}_{L} = Q_{\mathcal{G}_{L}} ( \tilde{\eta } \Psi )$ in $(\ref{definition of g})$ gives a generating function for vertices. 
Provided that the $t$-parametrization of $\Psi (t)$ is linear: $\Psi (t) = t \Psi$, we find 
\begin{align}
\label{linear t S} 
\sum_{n=1}^{\infty } \frac{\kappa ^{n-1} }{ (n+1)!}   \langle \Psi ,  V_{n} ( \Psi ^{n} ) \rangle  =\int_{0}^{1} {dt } \, \langle  \Psi , \eta \, \mathcal{G}_{L} ( t ) \rangle .
\end{align}
Note that all coefficients of $V_{n+1}$ and $\mathcal{G}_{L}^{(n)}$ match by the $t$-integral.

\vspace{3mm}

\underline{The generating function for $W_{n} ( \Psi ^{n} )$}

\vspace{2mm}

Let $\Psi _{\delta } = \sum_{n=0}^{\infty } \kappa ^{n} \Psi _{\delta } ^{(n)}$ be the expansion of the associated field $\Psi _{\delta }$ in powers of $\kappa$, where `$\delta$' is the variation operator. 
Recall that the variation of $S_{2}$ is given by $\delta S_{2} = \langle \delta \Psi , V_{1}(\Psi ) + W_{1} (\Psi ) \rangle =\langle \delta \Psi , \eta Q \tilde{\eta } \Psi \rangle $, which means $W_{1}(\Psi ) =0$.  
In section 2, we also determined $W_{2}$ and $W_{3}$, as well as $V_{2} $ and $V_{3}$, appearing in the calculation of the variation $\delta S_{3}$ and $\delta S_{4}$. 
Recall that $W_{2}$ is given by
\begin{align}
\frac{\kappa }{2} \langle \delta \Psi , W_{2} ( \Psi ^{2} ) \rangle   =  - \frac{\kappa }{2}  \langle  \delta \Psi  , [ \tilde{\eta } \Psi ,  V_{1} (\Psi ) ]^{L}  \rangle  ,
\end{align}
which is equivalent to $\langle \Psi _{\delta } ^{(1)} , V_{1} (\Psi ) \rangle $, and $W_{3}$ is given by 
\begin{align}
\nonumber 
\frac{\kappa ^{2}}{3!} \langle \delta \Psi ,W_{3} ( \Psi ^{3} ) \rangle & = - \frac{\kappa ^{2} }{3!} \langle  \delta \Psi ,  [ [ V_{1} ( \Psi ) , \tilde{\eta } \Psi ]^{L}  , \tilde{\eta } \Psi ]^{L} + 2 [ V_{1} (\Psi ) , Q \tilde{\eta } \Psi , \tilde{\eta } \Psi ]^{L} \rangle  
\\
& \hspace{20mm} + \frac{\kappa ^{2}}{2}  \langle \delta \Psi ,  \frac{1}{2} [ V_{2} (\Psi ^2) , \tilde{\eta } \Psi ]^{L} \rangle  ,
\end{align}
which is equivalent to $\langle \Psi _{\delta } ^{(1)} , \frac{\kappa }{2!}V_{2} ( \Psi ^{2} ) \rangle + \langle \Psi _{\delta } ^{(2)} , V_{1} ( \Psi ) \rangle $. 
Similarly, the following relation holds 
\begin{align}
\frac{\kappa ^{n}}{n!}  \langle \delta \Psi , W_{n} ( \Psi ^{n} ) \rangle = \sum_{k=1}^{n-1} \frac{\kappa ^{k-1}}{k!} \langle \Psi _{\delta } ^{(n-k)} ,  V_{k} ( \Psi ^{k } ) \rangle .
\end{align}
Hence, the associated field $\Psi _{\delta }$
\begin{align}
\Psi _{ \delta }  =  \delta \Psi  +  \frac{\kappa }{2} [ \tilde{\eta } \Psi  , \delta \Psi  ]^{L} 
+ \frac{\kappa ^{2}}{3!} \Big(  [ \tilde{\eta } \Psi  ,  Q \tilde{\eta } \Psi  , \delta \Psi  ] ^{L} 
 +2 \big[  \tilde{\eta } \Psi  , [ \tilde{\eta } \Psi  , \delta \Psi  ]^{L} \big] ^{L}  \Big) + \dots ,
\end{align}
defined by $\partial _{\tau } \Psi _{\delta } = \delta \Psi + \kappa [ \tilde{\eta } \Psi , \Psi _{\delta } ]^{L}_{\mathcal{G}_{L}}$ in $(\ref{large associated field})$ determines $W_{n}$-terms.

\vspace{3mm}

\underline{The WZW-like action}

\vspace{2mm}

Let $\Psi (t)$ be a $t$-parametrized NS-NS string field satisfying $\Psi (0) =0$ and $\Psi (1)=\Psi $. 
Replacing NS-NS string fields $\Psi $ with $t$-parametrized NS-NS string fields $\Psi (t)$ in (\ref{definition of g}) and (\ref{large associated field}), we obtain $t$-parametrized pure-gauge and large associated fields: $\mathcal{G}_{L}(t)$, $\Psi _{t}$, and $\Psi _{\eta }(t)$. 
WZW-like NS-NS action consists of these fields, which we explain in the rest. 
%We can extend the equation $(\ref{linear t S})$ to this $\Psi (t)$. 
%%%%%%%%%%%%%%%%%%%%
Since the relation
\begin{align}
\eta \, \mathcal{G}_{L} (t) = - Q_{\mathcal{G}_{L}(t)} \big( \tilde{\eta } \Psi _{\eta } (t) \big) = \tilde{\eta } \, Q_{\mathcal{G}_{L}(t)} \Psi _{\eta }(t)  
\end{align}
holds and $Q_{\mathcal{G}_{L}}$, $\eta $, and $\tilde{\eta }$ are nilpotent operators, the state $\eta \, \mathcal{G}_{L}$ is a $Q_{\mathcal{G}_{L}}$-, $\eta $-, and $\tilde{\eta }$-exact state. 
The equation $\ld \partial _{t} , \delta \rd \mathcal{G}_{L}(t) = 0$ implies $\partial _{t} (Q_{\mathcal{G}_{L}(t)} \Psi _{\delta }(t) ) = \delta ( Q_{\mathcal{G}_{L}(t)} \Psi _{t})$. 
%The $t$-parametrized associated field $\Psi _{t}$ is equivalent to $\Psi$ plus $\eta$-exact terms, provided that the $t$-parametrization is linear:$\Psi (t) = t \Psi$. 

\vspace{2mm}

Thus, we propose the following WZW-like action for NS-NS string field theory 
\begin{align}
S= \frac{2}{\alpha ^{\prime } } \int_{0}^{1} {dt} \, \langle  \eta \Psi _{t}  , \, \mathcal{G}_{L} (t)  \rangle ,
\end{align}
which reduces to $(\ref{linear t S})$ or the familiar WZW form (see Appendix B)
\begin{align}
S|_{\Psi (t) = t \Psi} = - \frac{1}{\alpha ^{\prime }} \Big( \langle  \Psi _{\eta } ,  \mathcal{G}_{L}  \rangle + \kappa  \int_{0}^{1} {dt}  \langle  \Psi  _{t} , [ \Psi _{\eta } (t)  , \mathcal{G} _{L} (t)  ]^{L}_{\mathcal{G}_{L} (t) } \rangle \Big) ,
\end{align}
if we set $\Psi (t) = t \Psi$. 
Note that the $(n+1)$-point vertex includes $n$ insertions of $\tilde{\eta }$ and the action $S$ is invariant under the linear\footnote{Note that, however, $\mathcal{G}_{L}$ and $\Psi _{t}$ include a lot of $\tilde{\eta } \Psi$ and as seen in 4.3, it does not mean that there are no gauge transformations including nonlinear terms of $\tilde{\eta } \Psi$. 
} gauge transformation $\delta _{\widetilde{\Omega }^{\prime} } \Psi  = \tilde{\eta } \, \widetilde{\Omega }^{\prime }$. 

\vspace{2mm}

The equation of motion is given by
\begin{align}
\label{e.o.m.}
\eta \, \mathcal{G}_{L} =   \int_{0}^{1} {d \tau } \Big(  \tilde{\eta } \eta \, Q_{\mathcal{G}_{L}[\tau ]} \Psi \Big)  =0,  
\end{align}
which is derived in subsection $4.3$. 
Although the action includes the integral over a real parameter $t$, the action $S$, the variation $\delta S$, the equation of motion $\eta \, \mathcal{G}_{L} =0$, and gauge transformations are independent of the $t$-parametrization or $t$-parametrized path $\Psi (t)$.

\subsection{Nonlinear gauge invariance}

Here, we derive the equation of motion and the closed form expression of nonlinear gauge transformations. 
Note that, for example, $\mathcal{G}_{L}(t=0) = 0$, $\mathcal{G}_{L}(t=1) = \mathcal{G}_{L}$, $\Psi _{\delta } (t=0 ) =0 $, and $\Psi _{\delta } (t= 1) = \Psi _{\delta }$ hold. 

\vspace{2mm}

For this purpose, we prove that the variation $\delta S$ does not includes $t$ and is given by
\begin{align} 
\delta S = \langle  \Psi _{\delta } ,  \eta  \, \mathcal{G}_{L}  \rangle .
\end{align}
Using the relation $\tilde{\eta } \, Q_{\mathcal{G}_{L}} ( \partial _{t} \Psi _{\delta } - \delta \Psi _{t} + \kappa [ \Psi _{t} , \psi _{\delta } ]^{L}_{\mathcal{G}_{L}}  ) =0$, which is equivalent to $\partial _{t} (\delta \mathcal{G}_{L} ) = \delta ( \partial _{t} \mathcal{G}_{L})$ with $\tilde{\eta } \Psi _{\mathbb{X} } = \psi _{\mathbb{X} }$, we find that the following equation holds for any $t$ %regardless of $t$-parametrization
\begin{align}
\nonumber 
\langle \delta \Psi _{t} , \, \eta \, \mathcal{G}_{L} (t) \rangle &= - \langle \delta \Psi _{t} , \, Q_{\mathcal{G}_{L} (t)} \psi _{\eta } (t) \rangle 
\\ \nonumber 
& = - \langle \partial _{t} \Psi _{\delta } (t) - \kappa [ \Psi _{\delta } (t) , \psi _{t} ]^{L}_{\mathcal{G}_{L} (t) }  ,  \, Q_{\mathcal{G}_{L} (t) } \psi _{\eta } ( t ) \rangle 
\\
&= \langle \partial _{t} \Psi _{\delta } (t) , \, \eta \, \mathcal{G}_{L} (t) \rangle - \kappa \langle \eta \, \mathcal{G}_{L} (t)  , \,  [ \Psi _{\delta } (t) , \psi _{t} ]^{L}_{\mathcal{G}_{L} (t) } \rangle .
\end{align}
Similarly, since $\eta \, \mathcal{G}_{L} =0$, $\ld \eta , Q_{\mathcal{G}_{L} } \rd =0$, and $\psi _{ \mathbb{X} } = \eta \, \Psi _{ \mathbb{X} }$, we obtain 
\begin{align}
\nonumber 
\langle \Psi _{t} , \, \delta \big( \eta \, \mathcal{G}_{L} (t) \big) \rangle &=  \langle \Psi _{t} , \, \eta \big( Q_{\mathcal{G}_{L} (t)} \psi _{\delta }(t) \big)  \rangle 
%= - \langle   Q_{\mathcal{G}_{L} (t)} ( \bar{ \eta }  \Psi _{t} )  , \, \psi _{\delta } (t)  \rangle 
=-\langle Q_{\mathcal{G}_{L} (t) } ( \eta  \Psi _{t } ) , \, \psi _{\delta }(t) \rangle 
\\ \nonumber 
&= - \langle   \psi _{\delta } (t) , \, Q_{\mathcal{G}_{L} (t) } (  \eta   \Psi _{t} ) \rangle 
= - \langle   \Psi _{\delta } (t) , \,   Q_{\mathcal{G}_{L} (t)} ( \eta   \psi _{t}  ) \rangle 
\\ \nonumber 
&=  \langle   \Psi _{\delta } (t) , \,  \eta  ( Q_{\mathcal{G}_{L} (t)}  \psi _{t} ) + \kappa  [ \eta \, \mathcal{G}_{L}(t) , \psi _{t} ] ^{L}_{\mathcal{G}_{L} (t)}  \rangle 
\\
&= \langle  \Psi _{\delta } (t) , \, \partial _{t} \big(  \eta \, \mathcal{G}_{L} (t) \big) \rangle + \kappa \langle \eta \, \mathcal{G}_{L} (t) , \,  [ \Psi _{\delta } (t) , \psi _{t} ]^{L}_{\mathcal{G}_{L} (t)} \rangle .
\end{align} 
Hence, the variation $\delta S$ of the WZW-like action $S$ is given by
\begin{align}
\nonumber 
\delta S =& \int_{0}^{1}{dt} \Big( \langle  \delta \Psi _{t} , \, \eta \, \mathcal{G}_{L} (t)  \rangle + \langle \Psi _{t} , \, \delta (\eta \, \mathcal{G}_{L} (t) ) \rangle \Big)   
\\
&= \int_{0}^{1}{dt} \, \frac{\partial }{\partial t} \, \langle \Psi _{\delta } (t) , \, \eta \, \mathcal{G}_{L} (t) \rangle 
= \langle \Psi _{\delta } , \, \eta \, \mathcal{G}_{L} \rangle ,
\end{align}
which does not include $t$-parametrized fields. 
The equation of motion is, therefore, given by $(\ref{e.o.m.})$ and it is independent of $t$-parametrization of fields. 

\vspace{2mm}

Since $\eta \, \mathcal{G}_{L} $ is a $Q_{\mathcal{G}_{L}}$-, $\eta$-, and $\tilde{\eta }$-exact state, we find that the action is invariant under the following nonlinear $Q$- and $\eta $-gauge transformations and linear $\tilde{\eta }$-gauge transformation 
\begin{align}
\Psi _{\delta } = Q_{\mathcal{G}_{L}} \Lambda + \eta \, \Omega  + \tilde{\eta } \, \widetilde{\Omega } , 
\end{align}
where $\Lambda$, $\Omega$, and $\widetilde{\Omega }$ are gauge parameter fields whose ghost-and-picture numbers are $(-1|0,0)$, $(-1| 1,0)$, and $(-1| 0,1)$ respectively.  
Note that $\Psi _{\delta }$ is an invertible function of $\delta \Psi$, at least in the expansion in powers of $\kappa $ as follows 
\begin{align}
\delta \Psi = \Psi _{\delta } - \frac{\kappa }{2} [ \tilde{\eta } \Psi , \Psi _{\delta } ]^{L} - \frac{\kappa ^{2}}{3!} \Big( \frac{1}{2} \big[ \tilde{\eta } \Psi , [ \tilde{\eta } \Psi , \Psi _{\delta } ]^{L} \big] ^{L} + 2 \big[ \tilde{\eta } \Psi , Q \tilde{\eta } \Psi , \Psi _{\delta } \big] ^{L} \Big) + O(\kappa ^{3}) . 
\end{align}
For instance, an explicit expression for $Q$-gauge transformation $\delta _{\Lambda } \Psi $ and $\eta $-gauge transformation $\delta _{\Omega } \Psi $ are given by
\begin{align}
& \hspace{20mm} \delta _{\Lambda } \Psi = Q \Lambda + \kappa [ Q \tilde{\eta } \Psi , \Lambda ]^{L} - \frac{\kappa }{2} [ \tilde{\eta } \Psi , Q \Lambda ]^{L} + O(\kappa ^{2} )
\\ 
\delta _{\Omega } \Psi &= \eta \Omega -  \frac{\kappa }{2} [ \tilde{\eta } \Psi , \eta \Omega ]^{L} - \frac{\kappa ^{2}}{3} \big[ \eta  \Omega , Q \tilde{\eta } \Psi , \tilde{\eta } \Psi  \big] ^{L} - \frac{\kappa ^{2}}{12} \big[  [ \eta \Omega , \tilde{\eta } \Psi  ] , \tilde{\eta } \Psi \big]  ^{L} + O(\kappa ^{3}) .
\end{align}
These gauge transformations are nonlinear. 
Note, however, that since $\tilde{\eta }$-gauge transformation 
\begin{align} 
\delta _{\widetilde{\Omega } } \Psi = \tilde{\eta } \widetilde{\Omega } - \frac{\kappa }{2} [ \tilde{\eta } \Psi , \tilde{\eta } \widetilde{\Omega} ]^{L} - \frac{\kappa ^{2}}{3}[ \tilde{\eta } \Psi , Q \tilde{\eta } \Psi , \tilde{\eta } \widetilde{\Omega } ]^{L} - \frac{\kappa ^{2}}{12} [ \tilde{\eta }\Psi , [ \tilde{\eta } \Psi , \tilde{\eta } \widetilde{\Omega } ]^{L} ]^{L}  + O(\kappa ^{3}) 
\end{align} 
obtained from $\Psi _{\delta _{\widetilde{\Omega }}} = \tilde{\eta } \, \widetilde{\Omega }$ consists of $\tilde{\eta }$-exact terms, 
it is equivalent to the linear $\tilde{\eta }$-gauge transformation 
\begin{align} 
\delta _{\widetilde{\Omega }^{\prime }} \Psi = \tilde{\eta } \, \widetilde{\Omega }^{\prime }, 
\end{align}
where $\widetilde{\Omega }^{\prime }$ is a redefined $\tilde{\eta }$-gauge parameter 
\begin{align}
\widetilde{\Omega }^{\prime } \equiv \widetilde{\Omega } - \frac{\kappa }{2} [ \tilde{\eta } \Psi , \widetilde{\Omega } ]^{L} - \frac{\kappa ^{2}}{3!} \Big( 2 \big[ \tilde{\eta } \Psi , Q \tilde{\eta } \Psi , \tilde{\eta } \widetilde{\Omega } \big] ^{L}  + \frac{1}{2} \big[ \tilde{\eta }\Psi , [ \tilde{\eta } \Psi , \tilde{\eta } \widetilde{\Omega } ]^{L}  \big] ^{L} \Big) + O(\kappa ^{3} ) . 
\end{align}
As a result, although the action has three generators of gauge transformations, since one of these gauge invariances reduces to trivial, the resulting theory is Wess-Zumino-Witten-likely formulated with two nonlinear gauge invariances.

\section{Conclusion}

In this paper, we proposed WZW-like expressions for the action and nonlinear gauge transformations in the NS-NS sector of superstring field theory in the large Hilbert space. 
Although the action uses $t$-parametrized large fields $\Psi (t)$ satisfying $\Psi (0)=0$ and $\Psi (1)=\Psi $, it does not depend on $t$-parametrization. 
%pure-gauge 
Vertices are determined by a pure-gauge solution of NS (heterotic) string field theory in the small Hilbert space of right movers, which is constructed by NS closed superstring products (except for the BRST operator) including insertions of right-moving picture-changing operators \cite{Erler:2014eba}. 

\vspace{2mm}

%gauge degree of large vertices
\underline{Gauge equivalent vertices} 

\vspace{1mm}

We used the $(-,{\rm NS})$ string products, namely, the right edge points at the diamonds of products in Figure 5.1 of \cite{Erler:2014eba}. 
It would be possible to write the large-space NS-NS action utilizing another but gauge-equivalent products in \cite{Erler:2014eba} instead of the $(-,{\rm NS})$ string products. 

\vspace{2mm}

%including R: 
\underline{Ramond sectors} 

\vspace{1mm}

We have not analyzed how to incorporate the R sector(s). 
Our large-space NS-NS action has the almost same algebraic properties as the large-space action for NS closed string field theory. 
Thus, we can expect that the method proposed in \cite{Kunitomo:2013mqa} also goes in the NS-NS case.

\vspace{2mm}

%It would be interesting if small-space formulation could be derived by partial gauge fixing of large-space theories
%geometric understanding, small/large correspondence, and gauge fixing
%Gauge fixing: It is in preparation. 
It is very important to obtain clear understandings of the geometrical meaning of theory, gauge fixing \cite{Torii:2011, Kroyter:2012ni}, the relation between two formulations: large- and small-space formulations. 
However, our large-space formulation is purely algebraic and these aspects remain mysterious. 
%Understand how large-space theories relate to formulations of superstring field theories in the small Hilbert space.

\vspace{5mm}

{\parindent=0pt{ {\bf{Acknowledgments:}} }}

\vspace{1mm}

The author would like to express his gratitude to the members of Komaba particle theory group, in particular, Keiyu Goto and my supervisors, Mitsuhiro Kato and Yuji Okawa. 
The author is also grateful to Shingo Torii. 
%%%%%%%%%%%%%%%%%%%%
This work was supported in part by Research Fellowships of the Japan Society for the Promotion of Science for Young Scientists

\vspace{5mm}

%\clearpage 

\appendix

\section{Heterotic theory in the small Hilbert  space}

The action for heterotic string field theory in the small Hilbert space of right movers is given by 
\begin{align}
S = \frac{1}{2} \langle  \Phi , Q \Phi \rangle + \sum_{n=1}^{\infty } \frac{\kappa ^{n}}{ (n+2)!} \langle  \Phi , [ \Phi ^{n} , \Phi ] ^{L} \rangle ,   
\end{align}
where the NS heterotic string field $\Phi$ is a ghost-and-picture number $(2|0 , -1)$ state in the small Hilbert space of right movers and right-moving picture-changing operators $\widetilde{X}$ inserted product $[ A_{1} , \dots , A_{n} ]^{L}$ given by \cite{Erler:2014eba} carries ghost-and-picture number $(3-2n | 0, n-1 )$. 
This action is invariant under the following gauge transformation \cite{Schubert:1991en, Zwiebach:1992ie} 
\begin{align}
\delta \Phi = Q \lambda  + \sum_{n=1}^{\infty } \frac{\kappa ^{n}}{n!} [ \Phi ^{n} , \lambda ]^{L} \equiv Q_{\Phi } \lambda , 
\end{align}
where $\lambda$ is a gauge parameter carrying ghost-and-picture number $(1|0,-1)$. 

Just as bosonic theory \cite{Sen:1990ff, Zwiebach:1992ie}, the equation of motion is given by 
\begin{align}
Q \Phi + \sum_{n=1}^{\infty }  \frac{\kappa ^{n}}{(n+1)!} [ \Phi ^{n} , \Phi ]^{L} =0 ,
\end{align}
and a pure-gauge $\mathcal{G}_{L}$ is constructed by infinitisimal gauge transformations \cite{Schubert:1991en, Berkovits:2004xh}. 
Therefore, $\mathcal{G}_{L}$ is defined by the $\tau =1$ value solution $\mathcal{G}_{L} \equiv \mathcal{G}_{L} [ \tau =1 ]$ of the following differential equation
\begin{align}
\frac{\partial }{\partial \tau } \mathcal{G}_{L} [\tau ] =  Q \lambda  + \sum_{n=1}^{\infty } \frac{\kappa ^{n}}{n!} \big[ {\mathcal{G}_{L}[\tau ] } ^{n} , \lambda \big] ^{L} = Q_{\mathcal{G}_{L} [\tau ] } \lambda , 
\end{align}
with the initial condition $\mathcal{G}_{L} [\tau =0] = 0$.

%\section{Some identities}

\section{Some identities}

%BPZ-properties are very useful when we compute some equations. 

\vspace{1mm}

\underline{BPZ-properties} 

\vspace{2mm}

The $c_{0}^{-}$-inserted BPZ inner product $\langle A , B \rangle := \langle \mathrm{bpz}(A)|c_{0}^{-} | B\rangle $, bosonic or heterotic string products, and a derivation operator $\mathbb{X}$ satisfy 
\begin{align}
\langle A , B \rangle & = (-)^{(A+1)(B+1)} \langle B , A \rangle ,
\\
\langle [ A_{0} , \dots , A_{n-1} ] , A_{n} \rangle & = (-)^{A_{0} + \dots + A_{n-1} } \langle A_{0} , [ A_{1} ,\dots , A_{n} ] \rangle ,
\\
\langle \mathbb{X} A , B \rangle & = (-)^{A \mathbb{X}} \langle A , \mathbb{X} B \rangle ,
\end{align}
where $c_{0}^{-} = \frac{1}{2} ( c_{0} - \tilde{c}_{0})$ and $\mathbb{X} = Q, \eta , \tilde{\eta }$.

\vspace{3mm}

\underline{The Maurer-Cartan element}

\vspace{2mm}

A pure-gauge solution $\mathcal{G}_{L}$ satisfies the equation of motion $\mathcal{F} ( \mathcal{G}_{L} )= 0$ of NS heterotic string field theory in the small Hilbert space of right movers. 
%Let $\mathcal{G}_{L} (t)$ be a $t$-parametrized one and consider 
Using the defining equation of $\mathcal{G}_{L}$, we find that 
\begin{align}
\nonumber 
\frac{\partial }{\partial \tau } \mathcal{F} ( \mathcal{G}_{L} ) & = \frac{\partial }{\partial \tau } \Big( Q \, \mathcal{G}_{L} + \sum_{n=1}^{\infty } \frac{\kappa ^{n}}{(n+1)!} \big[  \mathcal{G}_{L} ^{n} , \mathcal{G}_{L} \big] ^{L} \Big)
\\
&= Q Q_{\mathcal{G}_{L}} \tilde{\eta } \Psi + \sum _{n=1}^{\infty } \frac{\kappa ^{n}}{n!} \big[  \mathcal{G}_{L} ^{n} , Q_{\mathcal{G}_{L}} \tilde{\eta } \Psi \big] ^{L} 
= Q_{\mathcal{G}_{L}} ^{2} ( \tilde{\eta } \Psi ) ,
\end{align}
which leads to the differential equation $\partial _{\tau } \mathcal{F} = \big[  \mathcal{F} , \tilde{\eta } \Psi  \big] ^{L}_{\mathcal{G}_{L}}$ with the initial condition $\mathcal{F}(0)=0$. 
Hence, $\mathcal{G}_{L}$ satisfies $\mathcal{F}(\mathcal{G}_{L} ) =0$ and $Q_{\mathcal{G}_{L}} $ is a nilpotent operator. (See also \cite{Berkovits:2004xh}.)

\vspace{3mm}

\underline{The standard WZW form}

\vspace{2mm}

Recall that when there exist higher sting products $[ A_{1} , \dots , A_{n} ]^{L}$ $(n>2)$, a field-strength-like object $f_{XY} \equiv X \psi _{Y} - (-)^{XY} Y \psi _{X} + (-)^{X} \kappa [ \psi _{X} , \psi _{Y} ]^{L}_{\mathcal{G}_{L}}$ is not zero $f_{XY} \not= 0$ but a $Q_{\mathcal{G}_{L}}$-exact state: $Q_{\mathcal{G}_{L}} f_{XY}=0$, where $X$ and $Y$ are derivation operators satisfying $\ld X , Y \rd = 0$. 
Let 
\begin{align} 
F_{XY} \equiv  X \Psi _{Y} +  (-)^{(X+1)(Y+1)}Y \Psi _{X} + \kappa [ \Psi _{X} , \psi _{Y} ]^{L}_{\mathcal{G}_{L}} 
\end{align} 
be a {\it large} field-strength-like object satisfying $\tilde{\eta } F_{XY} = (-)^{X} f_{XY}$. 
Utilizing this $F_{\eta  t}$ and the relation $\langle \Psi _{t} , \, Q_{\mathcal{G}_{L} } \psi _{\eta } \rangle = \langle \Psi _{\eta } , \, \partial _{t} \mathcal{G}_{L} \rangle $, 
our WZW-like action can be rewritten as 
\begin{align}
\nonumber 
S &= \frac{1}{\alpha ^{\prime }} \int_{0}^{1} dt \, \Big( \langle \eta  \, \Psi _{t} , \, \mathcal{G}_{L} \rangle + \langle \Psi _{t} , \, \eta  \, \mathcal{G}_{L} \rangle \Big) 
\\ \nonumber 
&= \frac{1}{\alpha ^{\prime }} \int_{0}^{1} {dt } \Big(  \langle  F_{\eta  t} -  \partial _{t} \Psi _{\eta  } - \kappa [ \Psi _{t} , \psi _{\eta  }]^{L}_{\mathcal{G}_{L}} , \, \mathcal{G}_{L}  \rangle -  \langle \Psi _{t} , Q_{\mathcal{G}_{L}} \psi _{\eta } \rangle \Big) 
\\
& = \frac{1 }{\alpha ^{\prime }} \int_{0}^{1} {dt} \langle \mathcal{G}_{L} , \, F_{\eta  t} \rangle - \frac{1}{\alpha ^{\prime }} \int_{0}^{1} {dt } \Big[ \Big(  \langle  \partial _{t} \Psi _{\eta }, \mathcal{G}_{L} \rangle + \langle \Psi _{\eta } , \partial _{t} \mathcal{G}_{L} \rangle \Big) + \kappa \langle  \Psi _{t} ,  [ \psi _{\eta } , \mathcal{G}_{L} ]^{L}_{\mathcal{G}_{L}} \rangle \Big]   . 
\end{align}

Recall also that the linear $t$-parametrization $\Psi (t) = t \Psi $ gives $\Psi _{t} = \Psi $ up to $\tilde{\eta }$-exact terms. 
When we identify $\tau $ and $t$, the defining equation of $\psi _{X}$ becomes $\partial _{t} \psi _{X} = X \tilde{\eta } \Psi + \kappa[ \tilde{\eta } \Psi , \psi _{X} ]^{L}_{\mathcal{G}_{L}} $, which implies $\tilde{\eta } \big( \partial _{t} \Psi _{X} - (-)^{X} X \Psi + \kappa [ \Psi  , \psi _{X} ]^{L}_{\mathcal{G}_{L}} \big) =0$. 
Hence, provided that $\Psi (t) = t\Psi $, we obtain $\tilde{\eta } F_{\eta t}=0$ and the action reduces to the familiar WZW-form: 
\begin{align}
S|_{\Psi (t) = t \Psi} =  - \frac{1}{\alpha ^{\prime }} \Big( \langle  \Psi _{\eta } , \, \mathcal{G}_{L}  \rangle + \kappa  \int_{0}^{1} {dt} \, \langle  \Psi  _{t} , \, [ \psi _{\eta } (t)  , \mathcal{G} _{L} (t)  ]^{L}_{\mathcal{G}_{L}(t) } \rangle \Big) .
\end{align}


\small

\begin{thebibliography}{99}

%String fields

%\cite{Witten:1985cc}
\bibitem{Witten:1985cc}
  E.~Witten,
  ``Noncommutative Geometry and String Field Theory,''
  Nucl.\ Phys.\  B {\bf 268}, 253 (1986).
  %%CITATION = NUPHA,B268,253;%%

%\cite{Hata:1986}
\bibitem{Hata:1986}
  H.~Hata, K.~Itoh, T.~Kugo, H.~Kunitomo and K.~Ogawa,
  ``Covariant String Field Theory,''
  Phys.\ Rev.\  D {\bf 34} (1986) 2360, %. 
%  H.~Hata, K.~Itoh, T.~Kugo, H.~Kunitomo and K.~Ogawa,
  ``LOOP AMPLITUDES IN COVARIANT STRING FIELD THEORY,''
  Phys.\ Rev.\  D {\bf 35} (1987) 1356.
  %%CITATION = PHRVA,D35,1356;%%



%\cite{LeClair:1988}
%\bibitem{LeClair:1988}
%  A.~LeClair, M.~E.~Peskin and C.~R.~Preitschopf,
%  ``String Field Theory on the Conformal Plane. 1. Kinematical Principles, 2. Generalized Gluing,''
%  Nucl.\ Phys.\  B {\bf 317} (1989) 411, Nucl.\ Phys.\  B {\bf 317} (1989) 464.
  %%CITATION = NUPHA,B317,464;%%

%\cite{Kugo:1989}
%\bibitem{Kugo:1989}
%  T.~Kugo, H.~Kunitomo and K.~Suehiro,
%  ``Nonpolynomial Closed String Field Theory,''
%  Phys.\ Lett.\  B {\bf 226}, 48 (1989). 
%  T.~Kugo and K.~Suehiro,
%  ``NONPOLYNOMIAL CLOSED STRING FIELD THEORY: ACTION AND ITS GAUGE
%  INVARIANCE,''
%  Nucl.\ Phys.\  B {\bf 337}, 434 (1990).
  %%CITATION = NUPHA,B337,434;%%

%\cite{AlvarezGaume:1988bg}
\bibitem{AlvarezGaume:1988bg}
  L.~Alvarez-Gaume, C.~Gomez, G.~W.~Moore and C.~Vafa,
  ``Strings in the Operator Formalism,''
  Nucl.\ Phys.\ B {\bf 303} (1988) 455.
  %%CITATION = NUPHA,B303,455;%%
  %182 citations counted in INSPIRE as of 29 Jul 2014

%\cite{Sen:1990ff}
\bibitem{Sen:1990ff} 
  A.~Sen,
  ``Equations Of Motion In Nonpolynomial Closed String Field Theory And Conformal Invariance Of Two-dimensional Field Theories,'' 
   Phys.\ Lett.\ B {\bf 241}, 350 (1990).
  %%CITATION = PHLTA,B241,350;%%

%\cite{Schubert:1991en}
\bibitem{Schubert:1991en} 
  C.~Schubert,
  ``The Finite gauge transformations in closed string field theory,''  Lett.\ Math.\ Phys.\  {\bf 26}, 259 (1992).
  %%CITATION = LMPHD,26,259;%%

%\cite{Zwiebach:1992ie}
\bibitem{Zwiebach:1992ie}
  B.~Zwiebach,
  ``Closed string field theory: Quantum action and the B-V master equation,''
  Nucl.\ Phys.\  B {\bf 390}, 33 (1993)
  [arXiv:hep-th/9206084].
  %%CITATION = NUPHA,B390,33;%%

%\cite{Gaberdiel:1997ia}
\bibitem{Gaberdiel:1997ia}
  M.~R.~Gaberdiel and B.~Zwiebach,
  ``Tensor constructions of open string theories. 1: Foundations,''
  Nucl.\ Phys.\  B {\bf 505}, 569 (1997)
  [arXiv:hep-th/9705038].
  %%CITATION = NUPHA,B505,569;%%


%Superstring fields

%old-fashion

%\cite{Witten:1986qs}
\bibitem{Witten:1986qs}
  E.~Witten,
  ``Interacting Field Theory of Open Superstrings,''
  Nucl.\ Phys.\ B {\bf 276} (1986) 291.
  %%CITATION = NUPHA,B276,291;%%
  %449 citations counted in INSPIRE as of 29 Jul 2014

%\cite{Wendt:1987zh}
\bibitem{Wendt:1987zh}
  C.~Wendt,
  ``Scattering Amplitudes and Contact Interactions in Witten's Superstring Field Theory,''
  Nucl.\ Phys.\ B {\bf 314} (1989) 209.
  %%CITATION = NUPHA,B314,209;%%
  %83 citations counted in INSPIRE as of 29 Jul 2014

%\cite{Arefeva:1989cp}
\bibitem{Arefeva:1989cp}
  I.~Y.~Arefeva, P.~B.~Medvedev and A.~P.~Zubarev,
  ``New Representation For String Field Solves The Consistency Problem For Open Superstring Field Theory,''
  Nucl.\ Phys.\ B {\bf 341} (1990) 464.
  %%CITATION = NUPHA,B341,464;%%  %103 citations counted in INSPIRE as of 16 Mar 2013

%\cite{Preitschopf:1989fc}
\bibitem{Preitschopf:1989fc}
  C.~R.~Preitschopf, C.~B.~Thorn and S.~A.~Yost,
  ``Superstring Field Theory,''
  Nucl.\ Phys.\ B {\bf 337} (1990) 363.
  %%CITATION = NUPHA,B337,363;%%  %110 citations counted in INSPIRE as of 25 Mar 2013

%\cite{AlvarezGaume:1988sj}
\bibitem{AlvarezGaume:1988sj}
  L.~Alvarez-Gaume, C.~Gomez, P.~C.~Nelson, G.~Sierra and C.~Vafa,
  ``Fermionic Strings in the Operator Formalism,''
  Nucl.\ Phys.\ B {\bf 311} (1988) 333.
  %%CITATION = NUPHA,B311,333;%%
  %66 citations counted in INSPIRE as of 29 Jul 2014

%\cite{Saroja:1992vw}
\bibitem{Saroja:1992vw}
  R.~Saroja and A.~Sen,
  ``Picture changing operators in closed fermionic string field theory,''
  Phys.\ Lett.\ B {\bf 286} (1992) 256
  [hep-th/9202087].
  %%CITATION = HEP-TH/9202087;%%
  %14 citations counted in INSPIRE as of 26 Jul 2014

%\cite{Belopolsky:1997bg}
\bibitem{Belopolsky:1997bg}
  A.~Belopolsky,
  ``New geometrical approach to superstrings,''
  hep-th/9703183, 
  %%CITATION = HEP-TH/9703183;%%
  %19 citations counted in INSPIRE as of 29 Jul 2014
%
%\cite{Belopolsky:1997jz}
%\bibitem{Belopolsky:1997jz}
%  A.~Belopolsky,
  ``Picture changing operators in supergeometry and superstring theory,''
  hep-th/9706033.
  %%CITATION = HEP-TH/9706033;%%
  %16 citations counted in INSPIRE as of 29 Jul 2014

%\cite{Jurco:2013qra}
\bibitem{Jurco:2013qra}
  B.~Jurco and K.~Muenster,
  ``Type II Superstring Field Theory: Geometric Approach and Operadic Description,''
  JHEP {\bf 1304} (2013) 126
  [arXiv:1303.2323 [hep-th]].
  %%CITATION = ARXIV:1303.2323;%%
  %7 citations counted in INSPIRE as of 29 Jul 2014



%the large Hilbert space description 


%\cite{Berkovits:1995ab}
\bibitem{Berkovits:1995ab}
  N.~Berkovits,
  ``SuperPoincare invariant superstring field theory,''
  Nucl.\ Phys.\ B {\bf 450} (1995) 90
   [Erratum-ibid.\ B {\bf 459} (1996) 439]
  [hep-th/9503099].
  %%CITATION = HEP-TH/9503099;%%
  %184 citations counted in INSPIRE as of 26 Jul 2014

%\cite{Berkovits:1998bt}
\bibitem{Berkovits:1998bt}
  N.~Berkovits,
  ``A New approach to superstring field theory,''
  Fortsch.\ Phys.\  {\bf 48} (2000) 31
  [hep-th/9912121].
  %%CITATION = HEP-TH/9912121;%%
  %56 citations counted in INSPIRE as of 26 Jul 2014

%\cite{Berkovits:2004xh}
\bibitem{Berkovits:2004xh}
  N.~Berkovits, Y.~Okawa and B.~Zwiebach,
  ``WZW-like action for heterotic string field theory,''
  JHEP {\bf 0411} (2004) 038
  [hep-th/0409018], %.
  %%CITATION = HEP-TH/0409018;%%
  %27 citations counted in INSPIRE as of 26 Jul 2014
%
%\cite{Okawa:2004ii}
%\bibitem{Okawa:2004ii}
  Y.~Okawa and B.~Zwiebach,
  ``Heterotic string field theory,''
  JHEP {\bf 0407} (2004) 042
  [hep-th/0406212].
  %%CITATION = HEP-TH/0406212;%%
  %21 citations counted in INSPIRE as of 26 Jul 2014

%\cite{Matsunaga:2013mba}
\bibitem{Matsunaga:2013mba}
  H.~Matsunaga,
  ``Construction of a Gauge-Invariant Action for Type II Superstring Field Theory,''
  arXiv:1305.3893 [hep-th]. (To be replaced.)
  %%CITATION = ARXIV:1305.3893;%%
  %6 citations counted in INSPIRE as of 26 Jul 2014



%R sector 
 
%\cite{Berkovits:2001im}
\bibitem{Berkovits:2001im}
  N.~Berkovits,
  ``The Ramond sector of open superstring field theory,''
  JHEP {\bf 0111} (2001) 047
  [hep-th/0109100].
  %%CITATION = HEP-TH/0109100;%%
  %45 citations counted in INSPIRE as of 26 Jul 2014

%\cite{Michishita:2004by}
\bibitem{Michishita:2004by}
  Y.~Michishita,
  ``A Covariant action with a constraint and Feynman rules for fermions in open superstring field theory,''
  JHEP {\bf 0501} (2005) 012
  [hep-th/0412215].
  %%CITATION = HEP-TH/0412215;%%
  %20 citations counted in INSPIRE as of 29 Jul 2014

%\cite{Kunitomo:2013mqa}
\bibitem{Kunitomo:2013mqa}
  H.~Kunitomo,
  ``The Ramond Sector of Heterotic String Field Theory,''
  PTEP {\bf 2014} (2014) 4,  043B01
  [arXiv:1312.7197 [hep-th]], 
  %%CITATION = ARXIV:1312.7197;%%
  %3 citations counted in INSPIRE as of 26 Jul 2014
%
%\cite{Kunitomo:2014hba}
%\bibitem{Kunitomo:2014hba}
%  H.~Kunitomo,
  ``First-Order Equations of Motion for Heterotic String Field Theory,''
  arXiv:1407.0801 [hep-th].
  %%CITATION = ARXIV:1407.0801;%

%Amp.

%\cite{Friedan:1985ge}
\bibitem{Friedan:1985ge}
  D.~Friedan, E.~J.~Martinec and S.~H.~Shenker,
  ``Conformal Invariance, Supersymmetry and String Theory,''
  Nucl.\ Phys.\ B {\bf 271} (1986) 93.
  %%CITATION = NUPHA,B271,93;%%
  %1557 citations counted in INSPIRE as of 29 Jul 2014

%\cite{Berkovits:1999bs}
\bibitem{Berkovits:1999bs}
  N.~Berkovits and C.~T.~Echevarria,
  ``Four point amplitude from open superstring field theory,''
  Phys.\ Lett.\ B {\bf 478} (2000) 343
  [hep-th/9912120].
  %%CITATION = HEP-TH/9912120;%%
  %36 citations counted in INSPIRE as of 29 Jul 2014

%\cite{Iimori:2013kha}
\bibitem{Iimori:2013kha}
  Y.~Iimori, T.~Noumi, Y.~Okawa and S.~Torii,
  ``From the Berkovits formulation to the Witten formulation in open superstring field theory,''
  JHEP {\bf 1403} (2014) 044
  [arXiv:1312.1677 [hep-th]].
  %%CITATION = ARXIV:1312.1677;%%
  %2 citations counted in INSPIRE as of 26 Jul 201



%Reference




%\cite{Erler:2013xta}
\bibitem{Erler:2013xta}
  T.~Erler, S.~Konopka and I.~Sachs,
  ``Resolving Witten`s superstring field theory,''
  JHEP {\bf 1404} (2014) 150
  [arXiv:1312.2948 [hep-th]].
  %%CITATION = ARXIV:1312.2948;%%
  %3 citations counted in INSPIRE as of 26 Jul 2014
   
%\cite{Erler:2014eba}
\bibitem{Erler:2014eba}
  T.~Erler, S.~Konopka and I.~Sachs,
  ``NS-NS Sector of Closed Superstring Field Theory,''
  arXiv:1403.0940 [hep-th].
  %%CITATION = ARXIV:1403.0940;%%
  %2 citations counted in INSPIRE as of 26 Jul 2014






%Homotopy Algebras

%\cite{Stasheff:1963}
%\bibitem{Stasheff:1963}
%  J.Stasheff, 
%  ``Homotopy associativity of H-spaces I,II,''  Trans.\ Amer.\ Math.\ Soc. {\bf vol.108}(1963), 275-292, 293-312
%CITATION

%\cite{Getzler:1990}
%\bibitem{Getzler:1990}
%  E.Getzler, J.D.S.Jones, 
%  ``$A_{\infty }$-algebra and the cyclic bar complex,''  Journ. Math. {\bf 34}, 256(1990)
%CITATIOM

%\cite{Stasheff:1993ny}
\bibitem{Stasheff:1993ny} 
  J.~Stasheff,
  ``Closed string field theory, strong homotopy Lie algebras and the operad actions of moduli space,''  In *Penner, R. (ed.): Perspectives in mathematical physics* 265-288.  [hep-th/9304061].
  %%CITATION = HEP-TH/9304061;%%

%\cite{Kimura:1993ea}
\bibitem{Kimura:1993ea} 
  T.~Kimura, J.~Stasheff and A.~A.~Voronov,
  ``On operad structures of moduli spaces and string theory,''  Commun.\ Math.\ Phys.\  {\bf 171}, 1 (1995)  [hep-th/9307114].
  %%CITATION = HEP-TH/9307114;%%


%\cite{Kontsevich:1997vb}
%\bibitem{Kontsevich:1997vb} 
%  M.~Kontsevich,
%  ``Deformation quantization of Poisson manifolds. 1.,''  Lett.\ Math.\ Phys.\  {\bf 66}, 157 (2003)  [arXiv:q-alg/9709040 [q-alg]].  
  %%CITATION = ARXIV:Q-ALG/9709040;%%

%%\cite{Getzler:2007}
\bibitem{Getzler:2007}
  E.~Getzler,
  ``LIE THEORY FOR NILPOTENT $L_{\infty }$-ALGEBRAS''
  [arXiv:math/0404003].
  %%



  


%seeking for gauge fixing
  
  %\cite{Torii:2011zz}
\bibitem{Torii:2011}
  S.~Torii,
  ``Gauge fixing of open superstring field theory in the Berkovits non-polynomial formulation,''
  Prog.\ Theor.\ Phys.\ Suppl.\  {\bf 188} (2011) 272
  [arXiv:1201.1763 [hep-th]], 
  %%CITATION = ARXIV:1201.1763;%%
  %7 citations counted in INSPIRE as of 26 Jul 2014
%\cite{Torii:2012nj}
%\bibitem{Torii:2012nj}
  %S.~Torii,
  ``Validity of Gauge-Fixing Conditions and the Structure of Propagators in Open Superstring Field Theory,''
  JHEP {\bf 1204} (2012) 050
  [arXiv:1201.1762 [hep-th]]. %.
  %%CITATION = ARXIV:1201.1762;%%
  %8 citations counted in INSPIRE as of 26 Jul 2014

%\cite{Kroyter:2012ni}
\bibitem{Kroyter:2012ni}
  M.~Kroyter, Y.~Okawa, M.~Schnabl, S.~Torii and B.~Zwiebach,
  ``Open superstring field theory I: gauge fixing, ghost structure, and propagator,''
  JHEP {\bf 1203} (2012) 030
  [arXiv:1201.1761 [hep-th]].
  %%CITATION = ARXIV:1201.1761;%%
  %14 citations counted in INSPIRE as of 26 Jul 2014



\end{thebibliography}
\end{document}